\documentclass{aa}
\usepackage{graphics}
\input epsf

\def\r14{$R^{1 \over 4}$}
\def\hkpc{h^{-1}\,{\rm kpc}\,}

\def\kms{\,{\rm km~s^{-1}}}

\def\Aa{IR 09427+1929\,}
\def\Ab{IR 17432$-$5157\,}
\def\Ac{IR 23140+0348\,}
\def\Ad{IR 04025$-$8303\,}

\def\Ba{IR 00509+1225\,}
\def\Bb{IR 00276$-$2859\,}
\def\Bc{IR 02055+0835\,}
\def\Bd{IR 10026+4347\,}

\def\Ca{IR 14378$-$3651\,}
\def\Cb{IR 10559+3845\,}
\def\Cc{IR 09039+0503\,}
\def\Cd{IR 23242$-$0357\,}
\def\Ce{IR 20176$-$4756\,}

\def\beq{\begin{equation}}
\def\eeq{\end{equation}}

\def\Halpha{H$\alpha$}
\def\Hbeta{H$\beta$}

\begin{document}
\thesaurus{
	3
	(
11.16.1 
11.09.2 
11.05.1 
11.14.1 
11.19.1 
)}
\title{    An HST Surface Photometric Study of  
           Ultraluminous Infrared Galaxies
\thanks{This
research was based on observations obtained with the NASA/ESA {\it
Hubble Space Telescope} through the Space Telescope Science Institute,
which is operated by the Association of Universities for Research in
Astronomy, Inc., under NASA contract NAS5-26555.}
}
\author{Z. Zheng\inst{1,5}, H. Wu\inst{1,5}, S. Mao\inst{2},
	X.-Y. Xia\inst{3,5}, Z.-G. Deng\inst{4,5}, Z.-L Zou\inst{1,5},
}
\offprints{Z. Zheng (zhengz@bac.pku.edu.cn)\\ $^\dag$ BAC is jointly 
sponsored by the Chinese Academy of Sciences and Peking University.}
\institute{
 {$^{1}$} Beijing Astronomical Observatory, Chinese Academy of Sciences,
          100080 Beijing, P. R. China\\
 {$^{2}$} Max-Planck-Institut f\"ur Astrophysik
          Karl-Schwarzschchild-Strasse 1, 85740 Garching, Germany \\
 {$^{3}$} Dept. of Physics, Tianjin Normal University, 300074
          Tianjin, P. R. China\\
 {$^{4}$} Dept. of Physics, Graduate School, Chinese Academy of Sciences,
          100039 Beijing, P. R. China\\ 
 {$^{5}$} Beijing Astrophysics Center (BAC)$^\dag$, 100871,
        Beijing, P. R. China
 }
\date{Received ......, 1998; Accepted ......, 1998}

\maketitle

\markboth{An HST Surface Photometric Study of Ultraluminous Infrared Galaxies}{}

\begin{abstract}
We study the surface photometry for 13 single-nucleus
ultraluminous IRAS galaxies (ULIRGs),
selected from a parent sample of 58 
galaxies obtained in a Hubble Space Telescope snapshot survey.
We find that these galaxies can be classified into three classes
according to their surface photometry.
The surface brightness profiles of the four objects in the
first class are well fitted by the \r14 law.
The isophotes are all disky at $R \la 1\hkpc$, consistent with the 
molecular disks/rings found in nearby ULIRGs from CO observations. 
Each of the four galaxies in the second class has an inner
\r14 component plus an outer extension. 
Remarkably all these four galaxies are Seyfert 1
galaxies with luminosities in the quasar regime and with relatively
narrow permitted and strong FeII emission lines. The 
remaining five galaxies fall into the third class; these objects have
surface brightness profiles that deviate to various degrees from the \r14
law; indeed, one is fitted perfectly by an exponential law. We also present new
spectroscopic observations for some of these galaxies. We establish
the redshift of \Aa\, to be 0.284, instead of 0.149 as adopted 
in the literature. These observations firmly support the idea
that merging of disk galaxies produce
ellipticals. Our data also strongly suggest that the formation of
QSOs may be an integral part of 
elliptical galaxy formation (at low redshifts).
\end{abstract}

\section{Introduction}

Ultraluminous IRAS galaxies (ULIRGs), characterized by their high
far-infrared luminosities $L_{\rm IR} > 10^{12}L_{\odot}$
(for $H_0=50~{\rm km~s^{-1}~Mpc^{-1}}$), are the
most luminous galaxies in the local universe (see Sanders \& Mirabel 1996 
for a review). Most ULIRGs show signs of interaction and/or merging.
These processes are believed to 
trigger starbursts and central nuclear activities which in turn produce
intense infrared emissions via dust absorption and re-emission. Some
ULIRGs do not show prominent interaction signatures
(at least from ground-based imaging),  but these could be
advanced mergers, as is the case for the ULIRGs with highest luminosities
($L_{\rm IR} > 2.25\times 10^{12}L_{\odot}$, for $H_0=50~{\rm km~s^{-1}~Mpc
^{-1}}$). These ULIRGs may represent an important stage in formation of 
QSOs and powerful radio galaxies, and may also present 
a primary stage in the formation of elliptical galaxies.
(Sanders et al 1988a; Sanders \& Mirabel 1996; Melnick \& Mirabel 1990). 

Although the ultimate fate of the ULIRGs is not yet fully understood, 
more and more observations show that the remnants of the 
mergers resemble elliptical galaxies or S0 galaxies. For example,
the surface brightness profile of IRAS 20551-4250 is reasonably well 
fitted by the \r14 law (Johansson 1991).
Arp 220 and NGC 6240, two nearest ULIRGs, both have
densities, velocity dispersions and central surface brightness
distributions consistent with the fundamental plane of elliptical
galaxies (Doyon et al 1994). Baker \& Clements (1997) investigated
the old stellar population for 10 nearby ULIRGs based on deep near-infrared
imaging and concluded that 8 out of 10 ULIRGs show signs of elliptical
structure.

On the theoretical side, Toomre \& Toomre (1972) first successfully
reproduced the observed bridges and tails using
restricted three-body simulations of two colliding disk galaxies. In 1977,
Toomre put forward the hypothesis that all elliptical galaxies could be 
remnants of merged disk galaxies. Later simulations improved the spatial
resolutions, explored more phase space of the encounter geometry and
incorporated more realistic treatments of the
progenitors, adding components such as bulges, massive dark matter haloes and
gas; crude treatment of star formation is also attempted
(e.g. White 1978; Farouki \& Shapiro 1982; 
Negroponte \& White 1983; Barnes 1992; Hernquist 1992;
Heyl et al 1994; Barnes 
\& Hernquist 1996; Weil \& Hernquist 1996; Walker et al 1996; Mihos \& 
Hernquist 1994). 
These simulations conclude that a wide range of initial conditions 
can result in the formation of early-type-like galaxies (see de Zeeuw \& 
Franx 1991 for a review). Besides the global elliptical-like characteristics 
of the remnants of simulated mergers (e.g. the \r14 law of the
surface brightness profile), some simulations also attempt to reveal the fine 
structure of the remnants such as the shape of the isophote, loops, shells
and tails (see e.g. Heyl et al 1994; Springel \& White 1998).
As mergers of galaxy pairs may
not account for all observed properties of ellipticals, multiple mergers
have been considered (Mamon 1987, Barnes 1984, Barnes 1985, Barnes 1989, 
Schweizer 1989; Weil \& Hernquist 1996); multiple merging
may occur in compact groups such as those found by Hickson (1982, 1993).

While the simulations indicate that merging of two or more disk galaxies
can easily lead to the formation of ellipticals, it is clearly important
to test these predictions empirically. For this purpose,
ULIRGs are ideal since they are closely related 
to galaxy merging and interactions (e.g., Murphy et al 1996, 
Clements et al 1996). 
In this paper, we analyze the surface photometry for 13 
ULIRGs observed with the Hubble Space Telescope (HST)
Wide Field Planetary Camera 2 (WFPC2),
and examine whether their profiles resemble
the \r14 law, as one would expect if they
ultimately form elliptical galaxies. The resolution of HST provides us a
powerful way to study the inner profiles and the fine 
structure details in these galaxies.
We will also address the important question whether the
surface photometry of these galaxies are associated with the central
active galactic nuclei (AGN) phenomenon. The structure of the paper is
as follows. In section 2, we describe how we select our ULIRG
sample. Data reduction including surface 
photometry is described in Section 3.
In Section 4, we classify these galaxies according to
their surface brightness profiles.
In Section 5, we discuss the spectroscopic properties of these
galaxies, using both newly obtained spectra and data from the
literature. In Section 6, we
summarize and discuss the implications of
our results. Throughout this paper we adopt an Einstein-de Sitter
universe ($\Omega_0=1$) and denote the Hubble constant as
$H_0=100h~{\rm km~s^{-1}~Mpc^{-1}}$. Since our objects are at
relatively low redshift ($z \la 0.35$), the assumption about the density
parameter ($\Omega_0$) is not critical, and $1\arcsec$ corresponds to
roughly $1\hkpc (z/0.1)$ for $z \la 0.4$.
   
\section{Sample Selection}

In this study, we use archive images from an HST snapshot survey 
of ultraluminous IRAS galaxies (Borne et al 1998). 
This snapshot survey was taken using the WFPC2 in the I bandpass (F814W).
The objects observed were mainly selected from the following samples: 
the bright samples of Sanders et al (1988a,b) and  
Melnick \& Mirabel (1990), and the QDOT sample (Leech et al 1994; Lawrence 
et al 1999). 
For each target in this survey, two 400s exposures were taken.
Most of target galaxies are centered on the 
Wide Field Camera chip 3 (800$\times$800 pixels, $0.0996''$ per pixel)
and a few in the Planetary Camera
chip (800$\times$800 pixels, $0.0455''$ per pixel).
For details of this survey, see Borne et al (1998). 
In total, there are 58 images available to
us from March 1996 to Jan. 1997 (except that \Bd\thinspace was taken in 
May 1997). The redshifts of these objects range from $0.04$ to $0.35$.

\begin{table}[ht]
\caption[]{Journal of Observations}
\begin{tabular}{llrc}
\\
\hline\noalign{\smallskip}
{\bf Target Name}  & {\bf RA}(2000) & {\bf Dec}(2000) & {\bf Date}\\
\noalign{\smallskip}
\hline
\noalign{\smallskip}
IR 09427+1929    & 09:45:29.1 &  19:15:50.0 & 23/05/96 \\
IR 17432$-$5157    & 17:47:09.9 & $-$51:58:44.0 & 27/10/96 \\
IR 23140+0348    & 23:16:35.2 &  04:05:17.0 &  03/09/96 \\
IR 04025$-$8303    & 03:57:11.3 & $-$82:55:16.0 & 02/07/96 \\
\noalign{\smallskip}
IR 00509+1225    & 00:53:34.8 &  12:41:36.1 & 29/08/96 \\ 
IR 00276$-$2859    & 00:30:04.1 & $-$28:42:25.6 &  05/10/96 \\
IR 02055+0835    & 02:08:06.7 &  08:50:03.6 &  05/10/96 \\
IR 10026+4347    & 10:05:41.9 &  43:32:39.4 &  01/05/97 \\
\noalign{\smallskip}
IR 14378$-$3651    & 14:40:59.4 & $-$37:04:33.2 & 02/09/96 \\
IR 10559+3845    & 10:58:39.3 &  38:29:06.4 & 19/06/96 \\
IR 20176$-$4756    & 20:21:11.1 & $-$47:47:07.1 & 28/09/96 \\
IR 09039+0503    & 09:06:34.1 &  04:51:28.3 & 17/11/96 \\
IR 23242$-$0357    & 23:26:50.2 & $-$03:41:05.5 & 24/08/96 \\
\hline
\end{tabular}
\end{table}

The near infrared observation (F814W) minimizes the effect of the absorption
by gas and dust in the target galaxies because the dust extinction 
at longer wavelengths is smaller.
Moreover, near infrared imaging is not sensitive to young
stellar populations. The I-band imaging is therefore more suitable
for our primary goal of understanding how the
old stellar populations relax dynamically during the merging process.

The ULIRGs in the Borne et al (1998) sample exhibit a variety of morphologies.
Some have only a single nucleus while others have two or multiple nuclei 
and display complex and disturbed configurations. These morphologies are 
likely connected to different stages of merging and interaction. 
In this paper, we will focus on studying properties of the late-products 
of merging and relaxation, therefore we choose a subsample of galaxies 
that are dominated by a single nucleus; these single-nucleus galaxies
are selected by checking the contour plot for each source. 
The targets in our sample all show merging features, such as tidal
tails, plumes, shells and other fine structures. The galaxies selected
through this way are probably all mergers at the late stage of 
interacting/merging (\S 4, see also
Wu et al 1998). In total, 13 targets are chosen out of
the 58 images. The journal of observations for these targets is 
listed in Table 1. 

\section{Data Reduction}

All observations were preprocessed ``on the fly'' through the standard Space
Telescope Science Institute pipeline as described by
Holtzman et al (1995). The images were 
calibrated (including bias subtraction and flat fielding)
with the most up-to-date version of the routine reference
files provided by the Institute at the time the images were taken.
After correction of hot pixels, our image processing mainly include
cosmic-ray removal and sky subtraction, which we describe below. 

\subsection{Cosmic-ray Cleaning}
  
With its high altitude, HST is 
free of atmospheric seeing effects, however this also means that
it is much more susceptible to cosmic ray hits than ground-based
telescopes. So it is essential to carefully remove the cosmic rays.

Since for each of our object there are two 400s exposures that are 
in general well aligned, the cosmic-ray removal is relatively
straightforward. We use the standard IRAF task ``combine'' with
a (conventional) $3\sigma$ clipping threshold to detect the cosmic-rays.
The cosmic-ray free images are then co-added.
The combined image has a higher signal-to-noise ratio than  
each single image,
usually by a factor of $\sqrt{2}$ except for some individual pixels.

\subsection{Sky Subtraction}

The background of each combined image was fitted by performing the 
IRAF task ``imsurfit''. Since
the field of view for the Wide Field Camera
is small (about $80''\times80''$), 
it is reasonable that the sky background has a relatively small 
variation across the field.
After several trials, we found that it is sufficient to fit the sky 
background using a two-dimensional second-order Legendre function,
i.e., a ``declined plane''. Some rectangular
regions (usually each includes more than 2000 pixels)
were chosen from the combined image to sample the sky background.
The median value of each region was used for the fitting. 

The sky-subtracted image is then ready for photometric purposes.
    
\subsection{Surface Photometry}

We perform the surface photometry
using the ISOPHOTE package in STSDAS in the IRAF environment.

Foreground stars and diffraction spikes caused by bright stars or
central nuclei in each image were carefully masked 
out to eliminate their effect on the 
photometry. We use the SExtractor galaxy photometry package (v1.2b10b)
(Bertin \& Arnouts 1996) to help identify the stars. A circular 
region and a triangular region are used to mask a star and a spike,
respectively. Pixels inside these regions are given zero weighting
in the ellipse fitting procedure which we describe below.

\subsubsection{Isophote Ellipse Fitting}

For all galaxies, the IRAF task ``ellipse'' was used to derive
the isophotal morphological parameters.

The ``ellipse'' task fits ellipses which best reproduce the 
observed isophotes of an image (Jedrzejewski 1987). The best parameters
are found by an iterative procedure. In each run of the 
isophote fitting, a number of parameters are used to describe the fitted
ellipse. These include
the ellipse center ($x_0, y_0$), semi-major axis $a$, ellipticity 
$\epsilon$ ($\epsilon \equiv 1-b/a$, where $b$ is the semi-minor axis),
position angle $\phi$ (the azimuthal angle of the major 
axis, measured counter-clockwise from the $+y$ direction in the image
here). To describe the deviation of isophotes from perfect ellipses, the
difference between the isophotal radius and the best ellipse fit
is usually expanded in Fourier series.
The coefficient of $\cos 4\theta$ term in the expansion, $B_4$, is given
particular importance in describing the deviation (e.g. Carter 1977, 
Jedrzejewski 1987; Kormendy \& Bender 1996): if $B_4$ is positive, 
the isophote is more elongated 
along the major axis than the best fitting ellipse and is said to be
{\it disky}; on the other hand, if $B_4$ is negative, 
the isophote will appear rectangular ({\it boxy}). The intensity 
around each isophote is measured by averaging the flux azimuthally. The 
ellipse center, ellipticity and position angle are allowed to change in
this work for two reasons: First,
ellipticity changes and isophote twisting (i.e. changes
of position angle) have already been noticed in elliptical
galaxies. Second, 
our sample mainly includes interacting/merging galaxies. Behavior of the 
changes in the parameters may reveal interesting structure details.

\begin{table*}[ht]
\caption[]{Properties for the Ultraluminous IRAS Galaxies}
\begin{tabular}{lcrcclcl}
\\
\hline\noalign{\smallskip}
{\bf Target Name} & {\bf Redshift} & {\bf M$_I$-$5\log h$} &
{\bf Log($h^2\cdot$L$_{fir}$/L$_\odot$)} &
{\bf R$_e$($h^{-1}$kpc)} & {\bf Spectral Type} &
{\bf Class} & {\bf Comments}\\
\noalign{\smallskip}
\hline
\noalign{\smallskip}
IR 09427+1929    & 0.28400 & $<$-23.21 &12.01 &2.75 & Sy1$^o$      & 1 &Strong FeII\\
IR 17432$-$5157    & 0.17500 &  -21.11 &11.73 &4.99 & LINER$^s$    & 1 &\\
IR 23140+0348    & 0.21980 &  -22.94 &11.79 &2.47 & LINER$^o$ & 1 & Radio Galaxy\\
IR 04025$-$8303    & 0.13660 & $<$-22.41 &11.55 &0.94 & Sy1$^l$      & 1 &\\
\noalign{\smallskip}
IR 00509+1225    & 0.06114 & $<$-22.72 &11.21 &2.69 & Sy1$^n$      & 2 & Strong FeII \\
IR 00276$-$2859    & 0.28000 & $<$-23.70 &12.04 &3.25 & Sy1$^n$      & 2 & Strong FeII\\
IR 02055+0835    & 0.34500 & $<$-23.31 &12.37 &0.80 & Sy1$^l$      & 2 & Strong FeII \\
IR 10026+4347    & 0.17800 & $<$-23.11 &11.59 &0.84 & Sy1$^l$      & 2 & Strong FeII \\
\noalign{\smallskip}
IR 14378$-$3651    & 0.06760 &  -21.17 &11.72 &- & LINER$^k$  & 3 &\\
IR 10559+3845    & 0.20660 &  -22.32 &11.74 &- & HII$^s$              & 3 &\\
IR 20176$-$4756    & 0.17810 &  -21.51 &11.81 &- & not AGN$^l$   & 3 &\\
IR 09039+0503    & 0.12500 &  -21.80 &11.70 &- & LINER$^o$   & 3 &\\
IR 23242$-$0357    & 0.18900 &  -21.64 &11.52 &- & HII$^o$              & 3 &\\
\hline
\end{tabular}
\bigskip
\begin{footnotesize}
$^n$ NED; $^s$ Spectra of QDOT;
$^l$ Lawrence et al 1999;
$^k$ Kim et al 1998;
$^o$ Our observation
\end{footnotesize}
\end{table*}

\subsubsection{Fitting the Surface Brightness Profile}

In order to address the questions mentioned in the introduction,
we checked the one-dimensional surface brightness 
profiles obtained through isophote ellipse fitting.

We fit the radial profiles with a S\'{e}rsic (1968) law (Marleau \& 
Simard 1998):
\beq
I(r)=I_e \exp(-b[({R \over R_e})^{1/n}-1]), ~ b = 1.9992n-0.3271,
\eeq
where $I(R)$ is the surface brightness at radius $R$,
and $R_e$ is the effective radius within which 
half of the light is enclosed (Capaccioli 1989).
The de Vaucouleurs \r14 profile, which is usually used to describe 
elliptical galaxies and bulges, is a special case 
of (1) with $n=4$, and the exponential 
profile, which is used to describe galactic disks,
is also a special case with $n=1$.

\section{Results}

Table 2 gives the properties for these galaxies, including their
redshifts, magnitudes, infrared luminosities and spectral classifications.
The integrated apparent WFPC2 magnitude is obtained from aperture 
photometry in the sky-subtracted and star-masked image using the formula 
from Holtzman et al (1995); the apparent magnitude is then converted into
absolute magnitude using luminosity distances for our cosmology (no
k-correction has been applied).
In the case of saturated images, only upper limits 
(in absolute magnitude) are given. Figs. 1-13 (left panels) show the HST 
images for all thirteen galaxies. Each image has $750 \times 750$ pixels, 
corresponding to $74.5\arcsec$ on a side. The middle panels show the contour 
levels for a smaller region centered on each target, highlighting its
environments. The right panels show
the variations of the surface brightness, ellipticity,
position angle and $B_4/a$ as a function of \r14. From the
radial surface brightness profiles we classify
these galaxies into three classes:

\begin{figure*}
\resizebox{5.50cm}{!}{\includegraphics{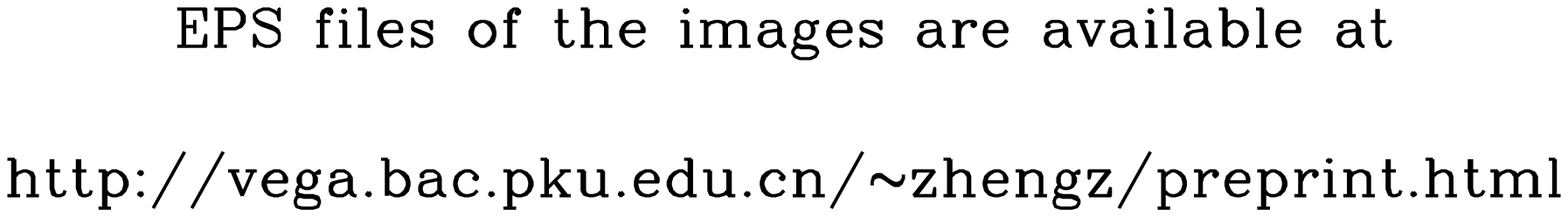}}
\resizebox{5.95cm}{!}{\includegraphics{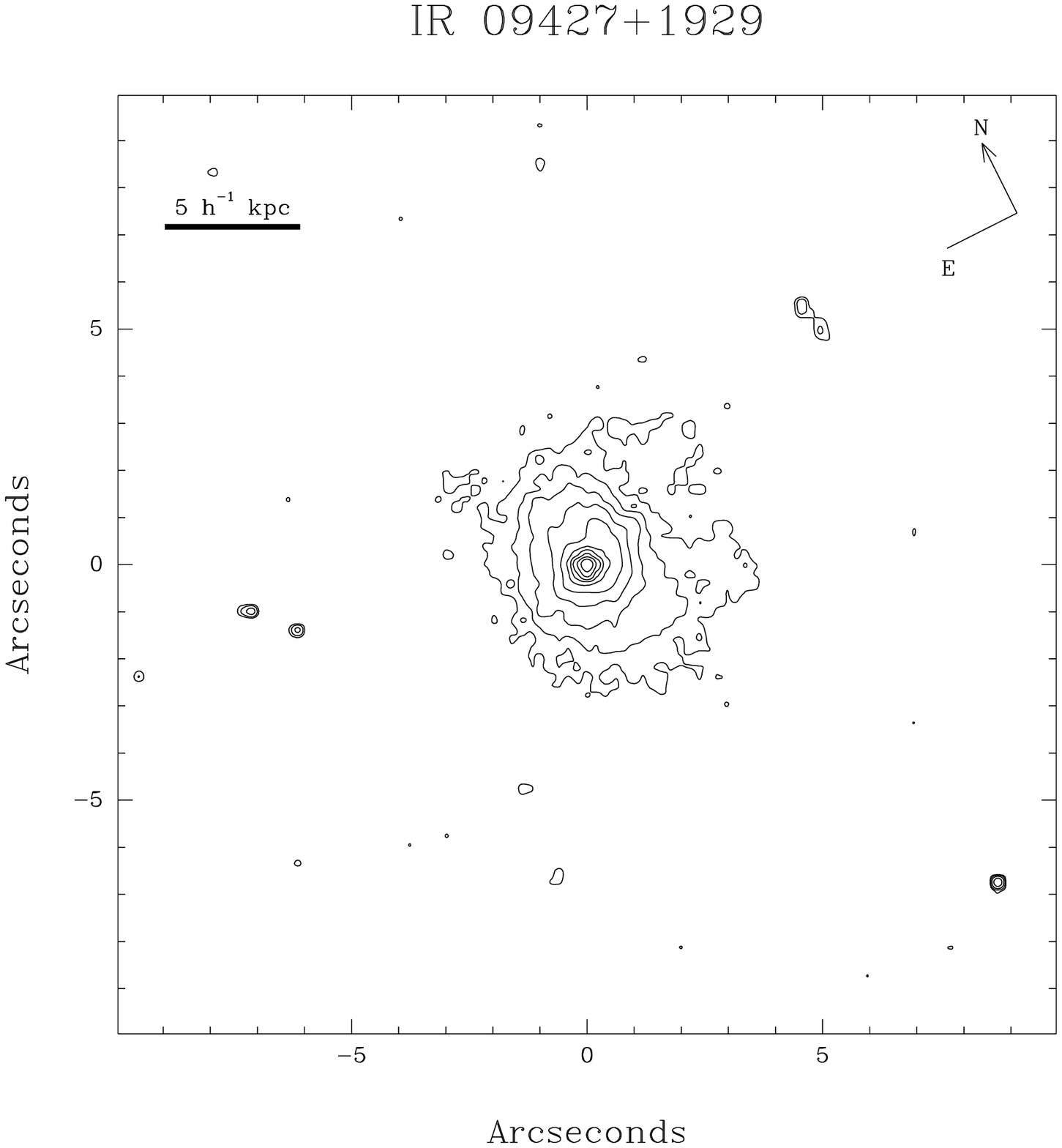}}
\resizebox{6.10cm}{!}{\includegraphics{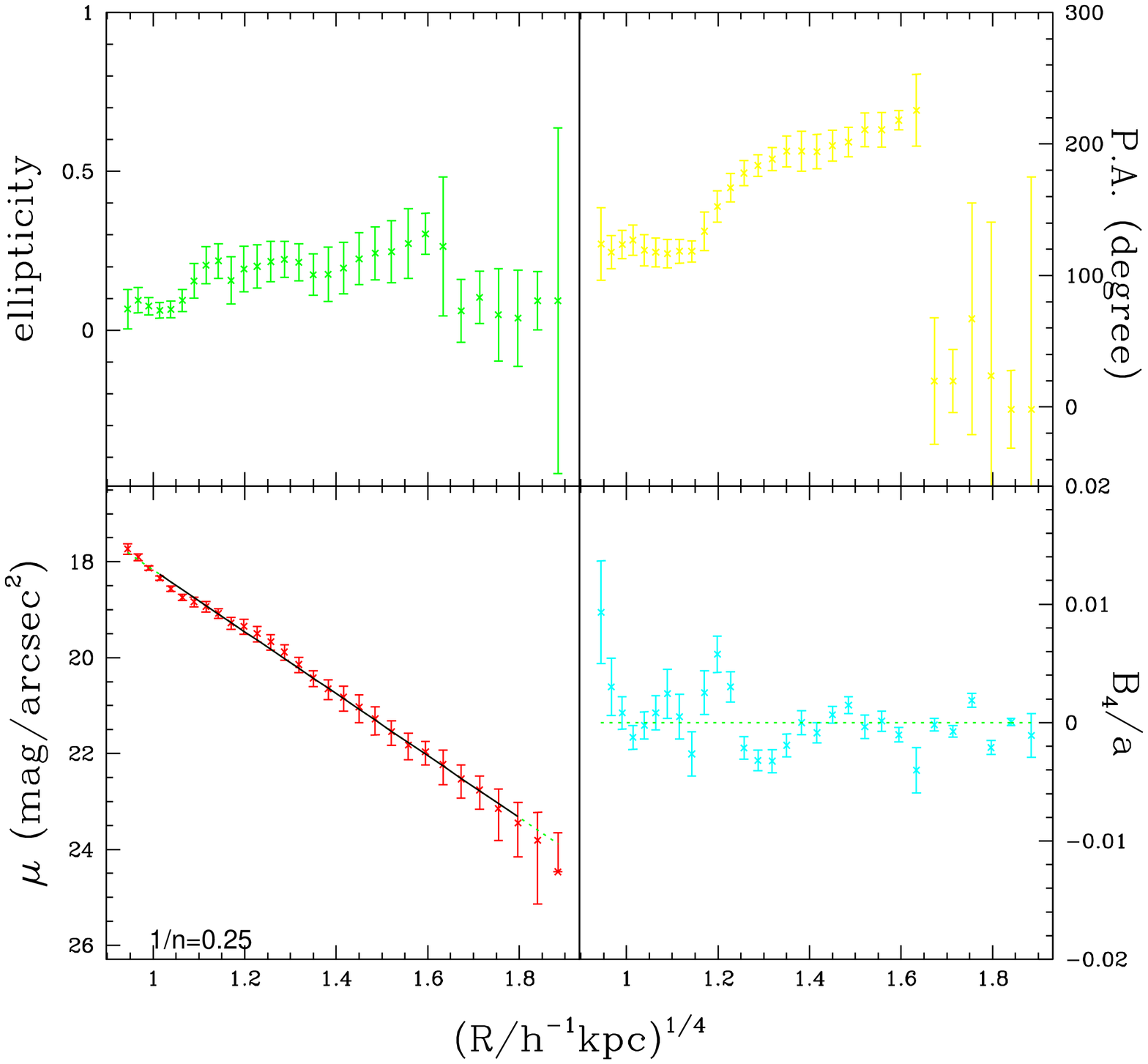}}
\caption{The left panel shows the HST Wide Field Camera image for
\Aa (the object labeled as G0).
The image shown is $74.5\arcsec$ on a side. Three additional
galaxies (G1, G2 and G3) are within the field. The ellipse shows the
error box of the IRAS position (cf. \S 5.1).
The middle panel shows the contours centered on \Aa; the contour 
levels are 2, 4, 8, 16, 32, 64, 128, 256, 512, 1024 ADU,
respectively. The thick
horizontal bar indicates a scale of 5$\hkpc$. The north and
east directions are indicated at the top right in the middle panel. The right
panels show the variations of the
ellipticity, position angle, 
surface brightness, and $B_4/a$ (see \S 3.3) as a function of \r14.
The best \r14 fit to the surface brightness profile
is given by the straight line.
}
\end{figure*}

\begin{figure*}
\resizebox{5.50cm}{!}{\includegraphics{url.ps}}
\resizebox{5.95cm}{!}{\includegraphics{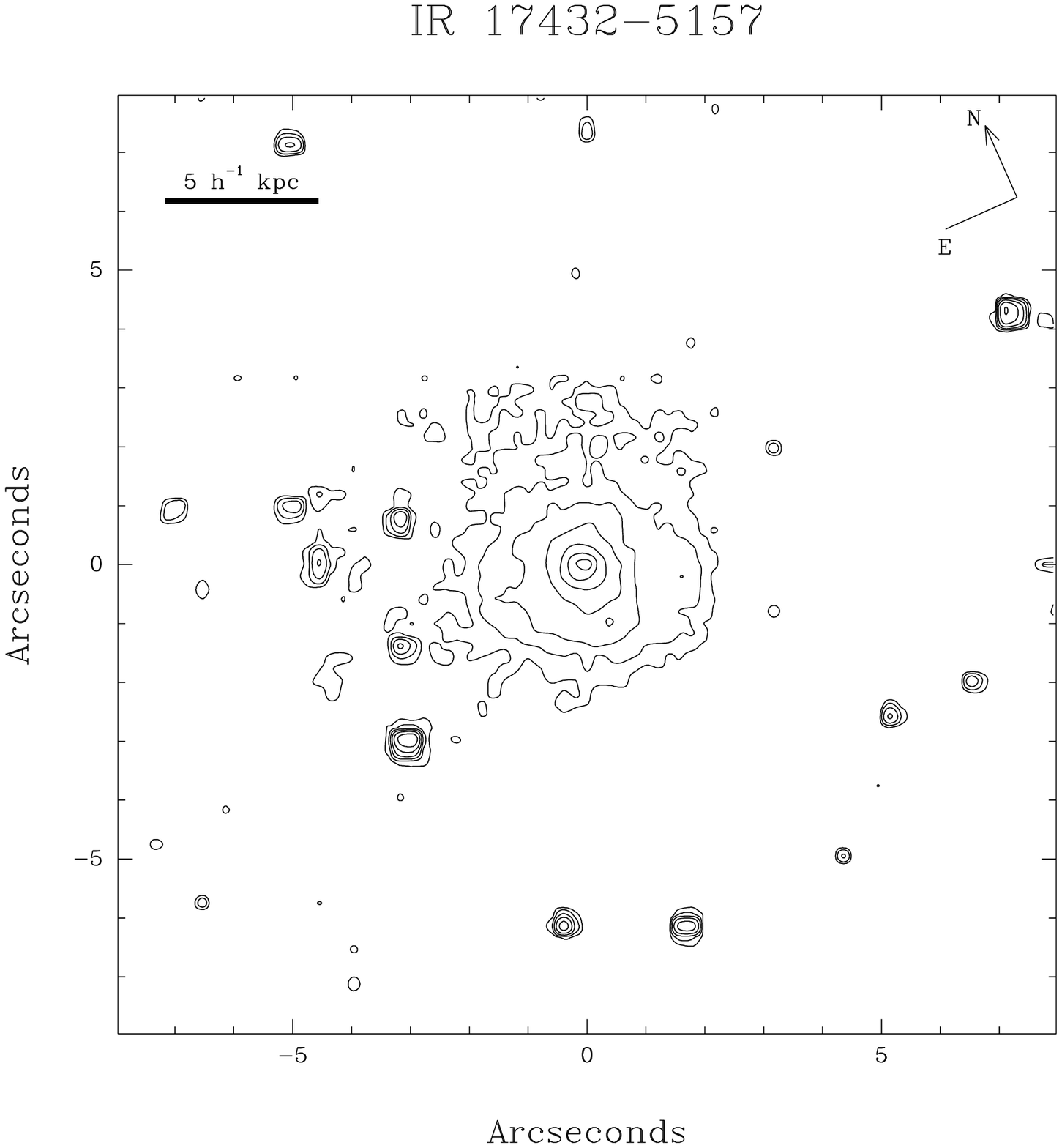}}
\resizebox{6.10cm}{!}{\includegraphics{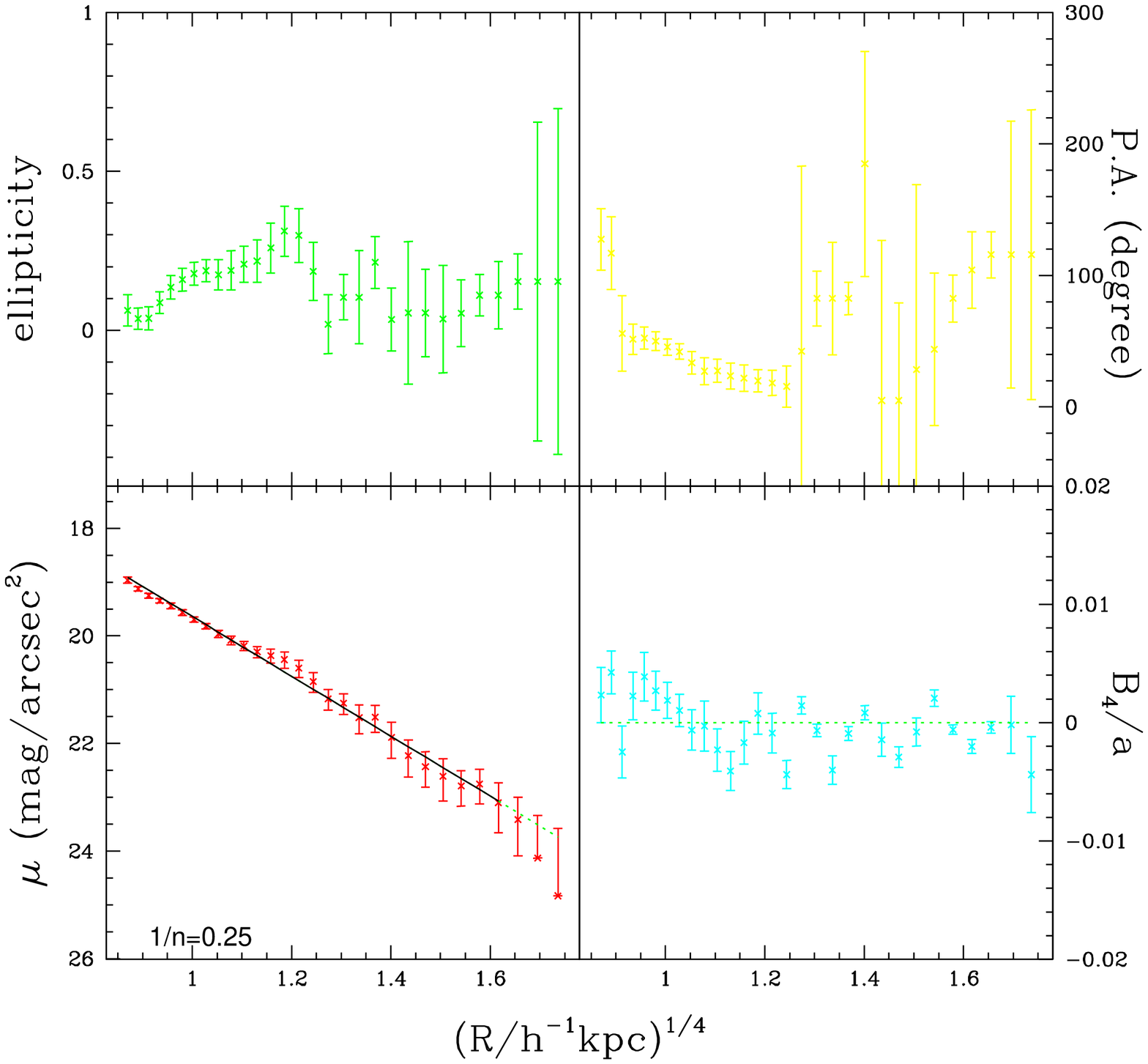}}
\caption{Same as in Fig. 1, for \Ab. At least one additional
galaxy are within the field. Due to its low Galactic latitude
($-12^\circ$), many stars are also present.
The contour levels are 
1.9, 3.8, 7.6, 15.2, 30.4, 60.8, 121.6, 243.2, 486.4 ADU, respectively.
}
\end{figure*}

\begin{itemize}
\item[I.] The first class includes 4 galaxies; the surface brightness
of each galaxy in this class is well fitted by the \r14
law with some small deviations at $\sim$ $\hkpc$.
These four galaxies have round appearances.
\item[II.] The second class includes 4 galaxies. 
The surface brightness profile for each galaxy in this class
has two components: an inner \r14 component and an outer
component. There is a bright central nucleus in each of these galaxies.
The best-fit effective radii (for the \r14 component) for class I and II 
galaxies are given in Table 2. 
\item[III.] The remaining 5 galaxies fall into the third class. 
These galaxies have surface brightness profiles that deviate significantly
from the \r14 law.
\end{itemize}

In the following, we will discuss the properties of these three 
classes in detail. 

\subsection{Class I}

Four galaxies 
({\bf \Aa, \Ab, \Ac, \Ad}) belong to class I. The surface brightness
profiles for these galaxies are well described by the
\r14 law, from the very inner part to the outer part as shown in Fig.
1 to Fig. 4. For \Ab, the fit
is reasonable from 0.5 $\hkpc$\,to $9.2 \hkpc$; 
for \Ac, the fit is excellent out to
$13 \hkpc$. Another common characteristic of these galaxies
is that they have round appearance.
For each galaxy, the {\it maximum} ellipticity 
of the isophotes is less than 0.30. The ellipticities
at the effective radius are even smaller. They are only 
0.2, 0.1, 0.1, and 0.1 for \Aa, \Ab, \Ac, \Ad, respectively.

\begin{figure*}
\resizebox{5.50cm}{!}{\includegraphics{url.ps}}
\resizebox{5.95cm}{!}{\includegraphics{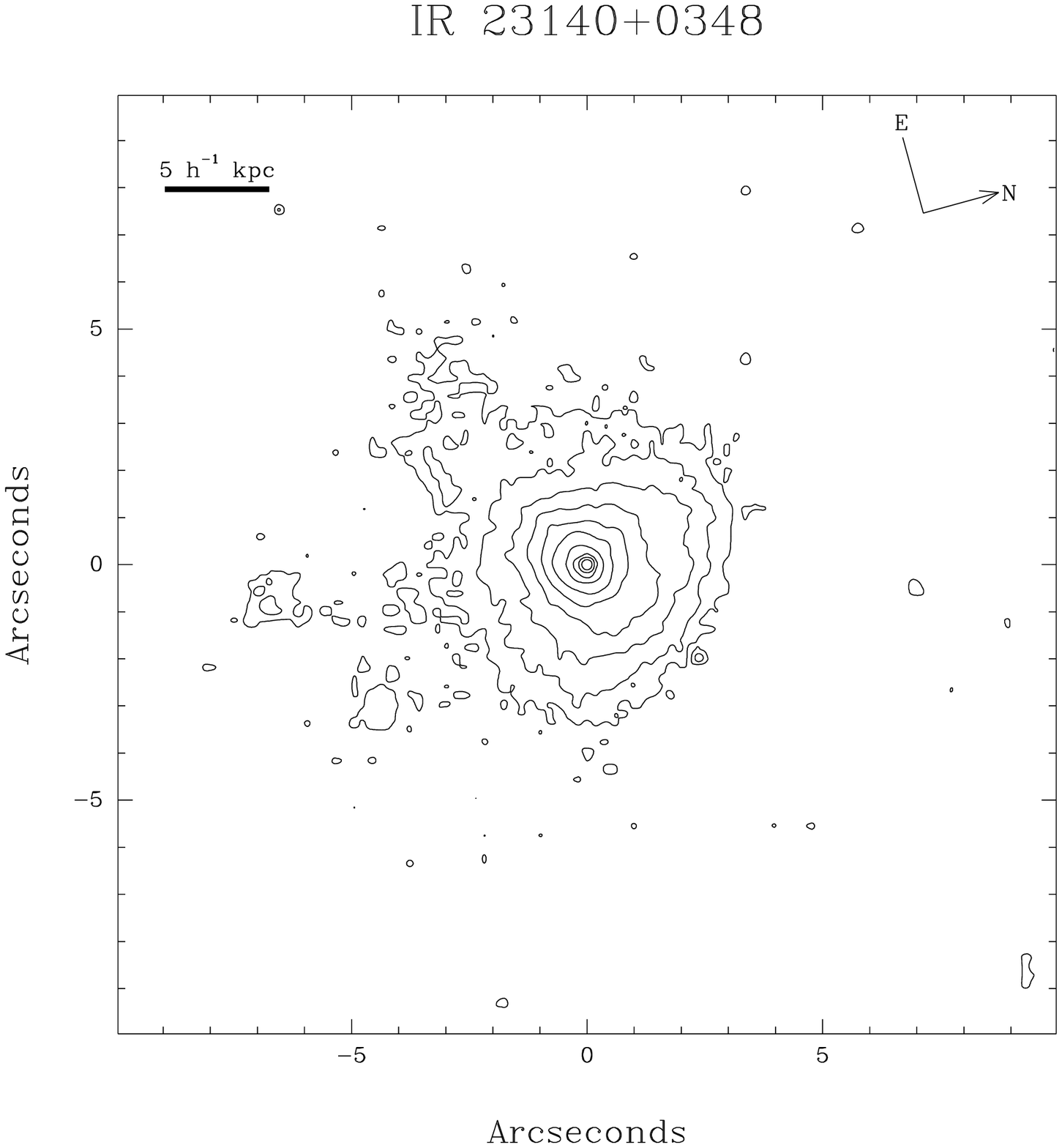}}
\resizebox{6.10cm}{!}{\includegraphics{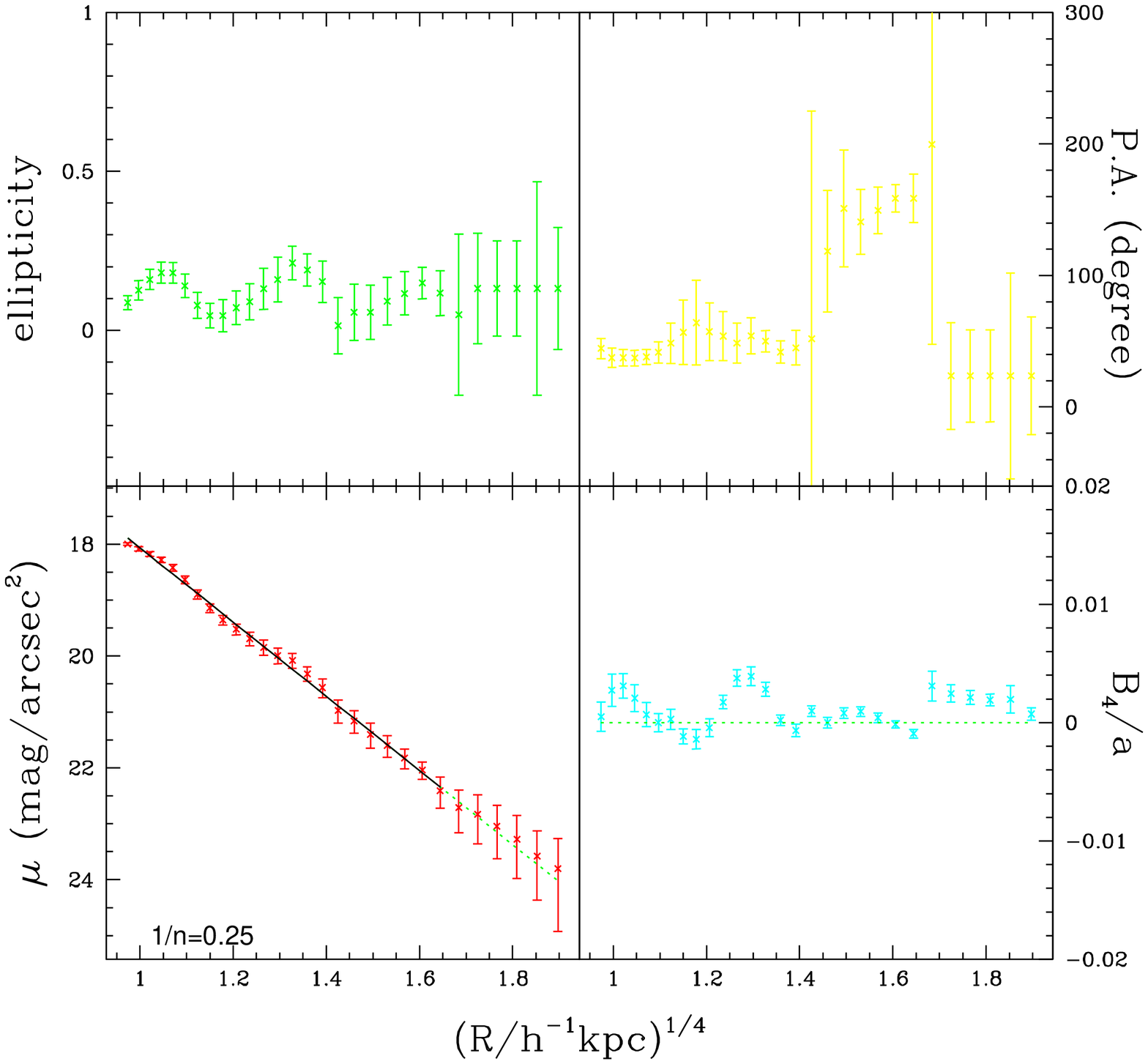}}
\caption{
Same as in Fig. 1, for \Ac (labeled as G0). Faint features are visible around 
G0. The contour levels are 
1.64, 3.28, 6.56, 13.12, 26.24, 52.48, 104.96, 209.92, 419.84, 839.68, 1679.36
ADU, respectively.}
\end{figure*}

\begin{figure*}
\resizebox{5.50cm}{!}{\includegraphics{url.ps}}
\resizebox{5.95cm}{!}{\includegraphics{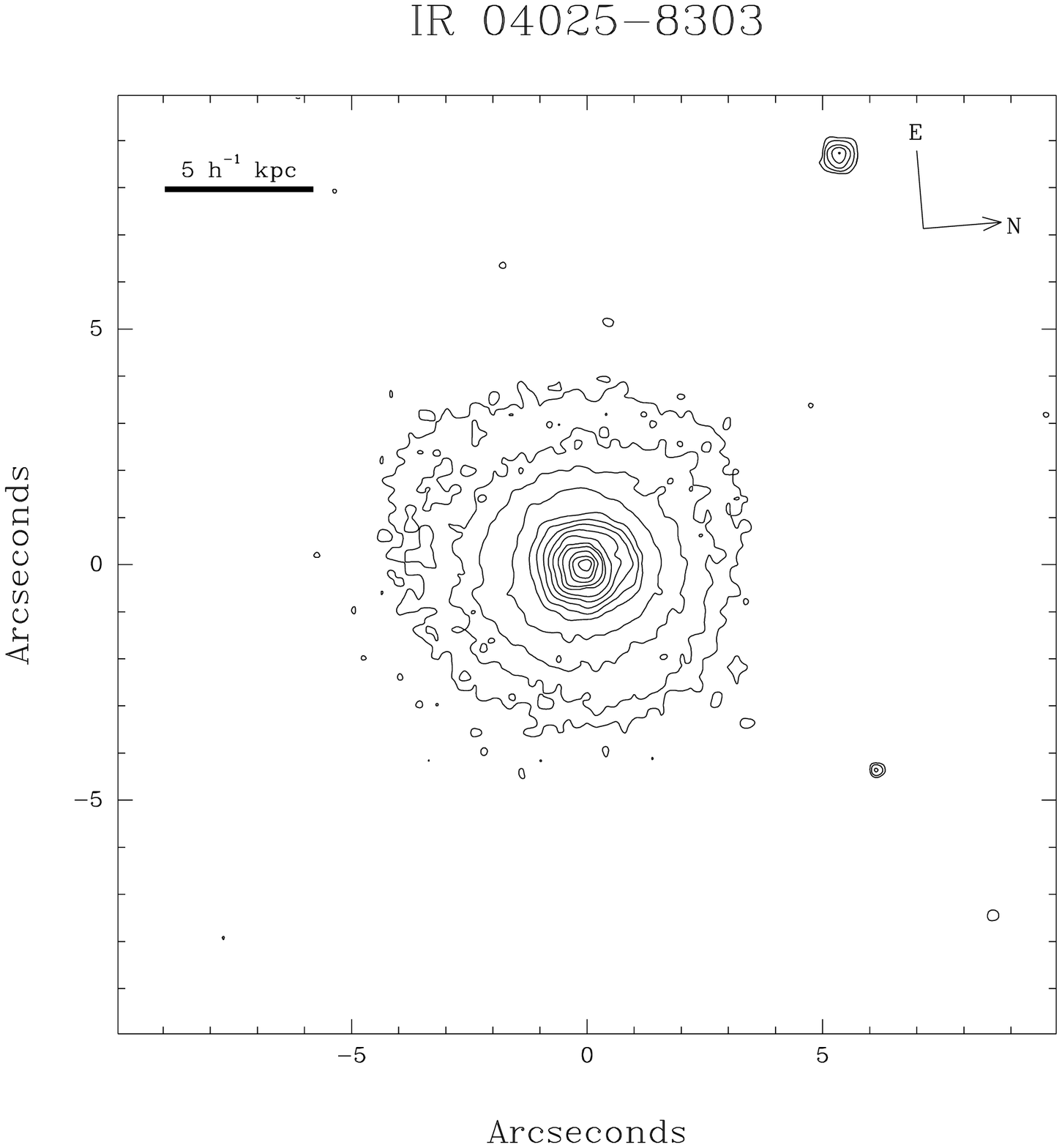}}
\resizebox{6.10cm}{!}{\includegraphics{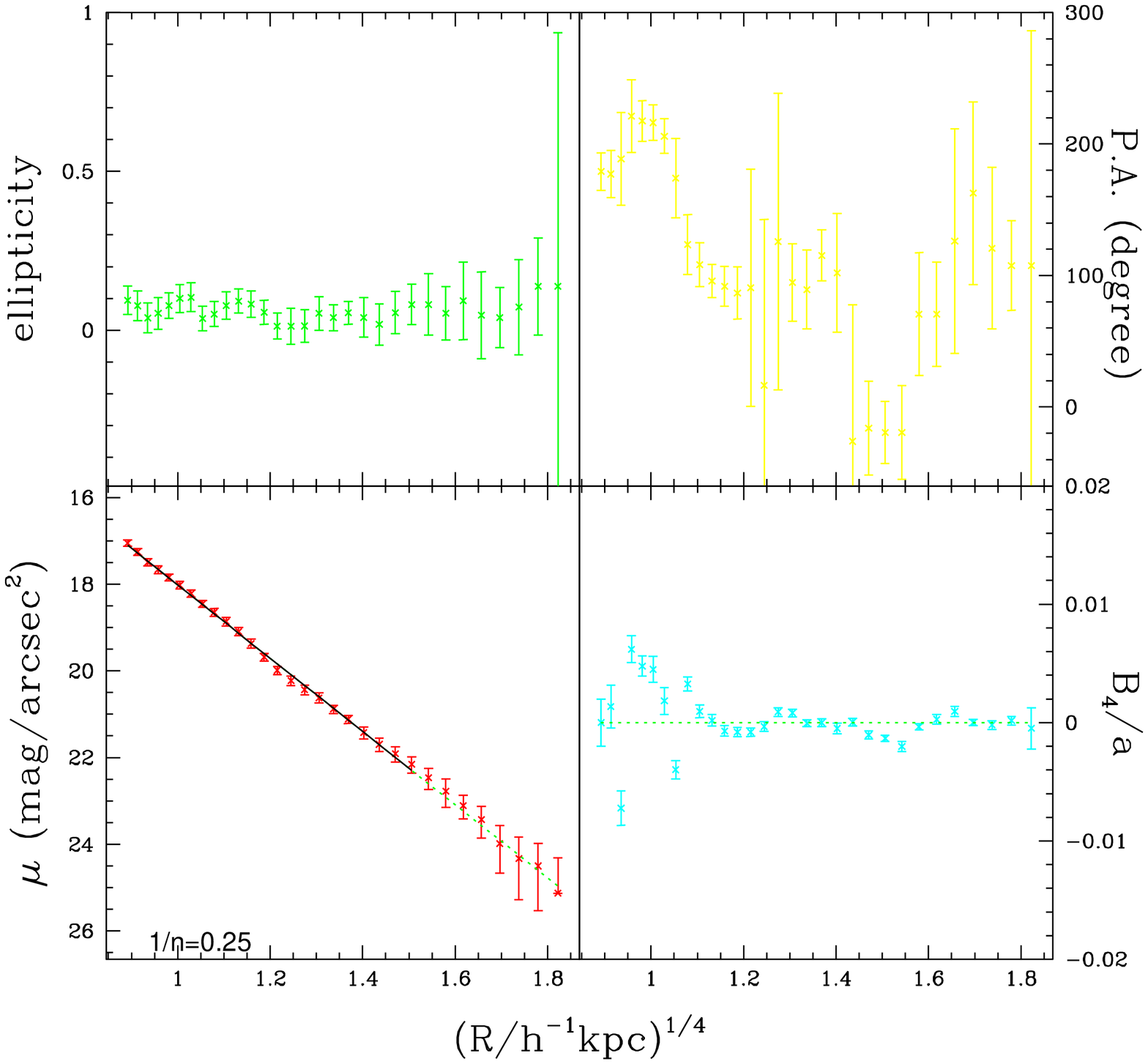}}
\caption{
Same as in Fig. 1, for \Ad. At least one additional
galaxy (G1) is within the field. The contour levels are 
1.55, 3.1, 6.2, 12.4, 24.8, 35, 49.6, 70, 99.2, 150, 198.4, 396.8, 
793.6, 1587.2, 3174.4 ADU, respectively.
}
\end{figure*}

However, there are slight deviations from the exact \r14 law.
Small excesses over the best-fitted curve 
can be seen in all surface brightness profiles (right panels in Figs. 1-4)
at $\sim (1-3) \hkpc$.
The bump position in each galaxy
coincides with the maximum of the ellipticity $\epsilon$.
This coincidence reflects small structures in these galaxies. Furthermore, 
it is very interesting that all four galaxies
have disky isophotes ($B_4>0$) in the very inner regions
($\leq 1h^{-1}$ kpc), and the position angle change is very large,
($\sim 100^\circ$ for some galaxies).
Carollo et al (1997) analysed HST WFPC2 images for
15 elliptical galaxies. In both F555W and F814W filters,
the position angle changes are usually less than $20^\circ$. Therefore
it appears that although the light profile is described by 
the \r14 law, the ellipses still show substantial twists.

   From Figs. 1-4, it can be seen that all four galaxies  
exhibit signatures of merging
such as short tails, plumes in their outer parts. Except for \Ac, all
the other galaxies have some other galaxies within the same field of
view, but these could be due to chance alignment (see Section 5).
The round appearance of these galaxies provides a clue to the
merging environment. Numerical simulations (e.g. Weil \& Hernquist 1996)
found that remnants of multiple mergers are more round
than the mergers from pair stellar disks
(when projected onto the sky). So the
round appearance of the class I galaxies suggests that they 
may be remnants of multiple mergers. Such a possibility
is supported by the high percentage of multiple nuclei galaxies
in our parent sample of 58 galaxies. About 50\% of these
have more than three nuclei. Furthermore, for
some of the double-nucleus galaxies, each of the
two merging nuclei could contain a pair of galactic nuclei in
high-resolution images, similar to Arp 220
(Taniguchi \& Shioya 1998). Therefore,
the percentage of the multiple-nucleus galaxies is
likely higher than 50\% for our sample, supporting the notion that
ULIRGs may preferentially reside in groups of galaxies (Wu et al 1998).

\begin{figure*}
\resizebox{5.50cm}{!}{\includegraphics{url.ps}}
\resizebox{5.95cm}{!}{\includegraphics{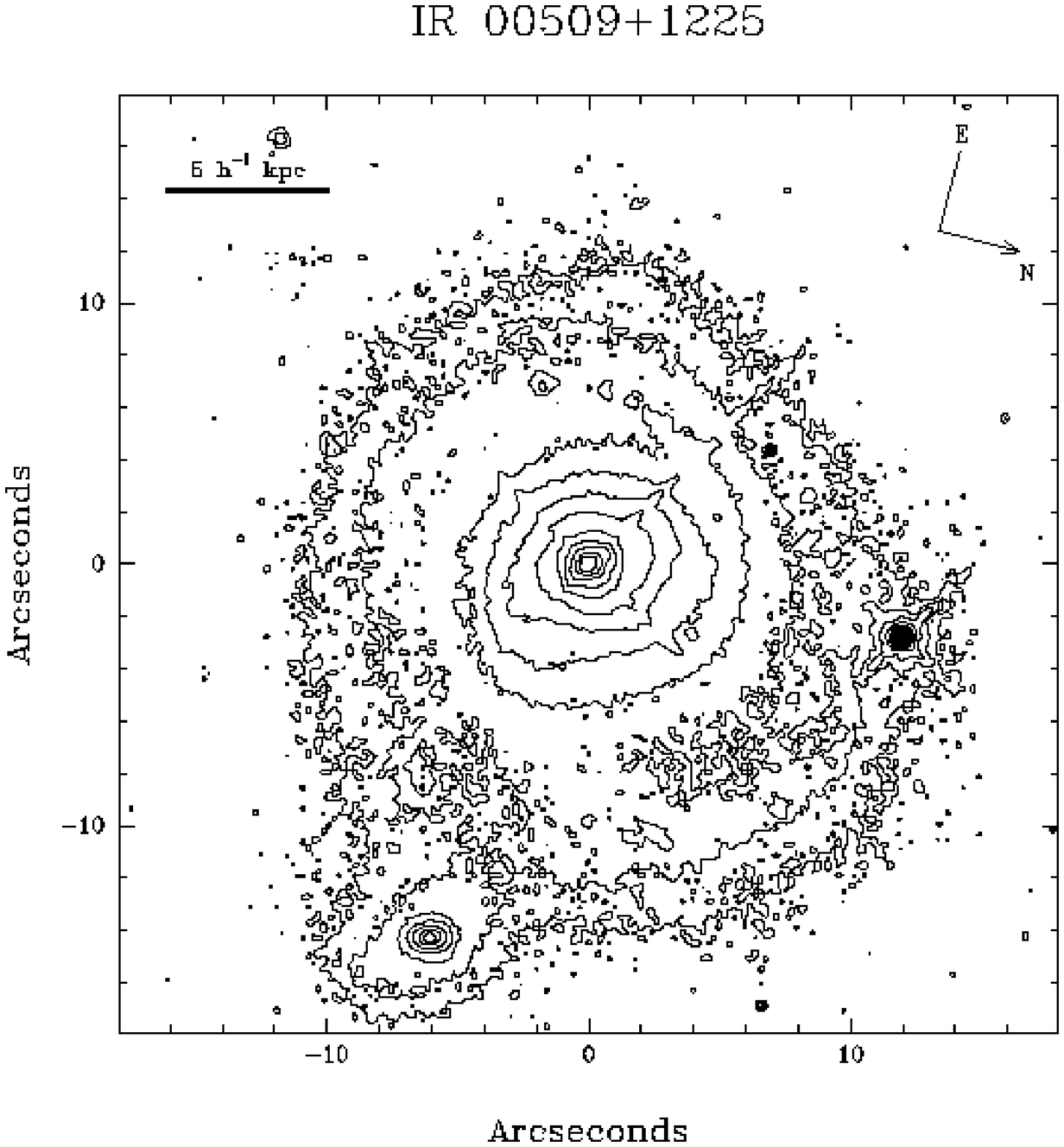}}
\resizebox{6.10cm}{!}{\includegraphics{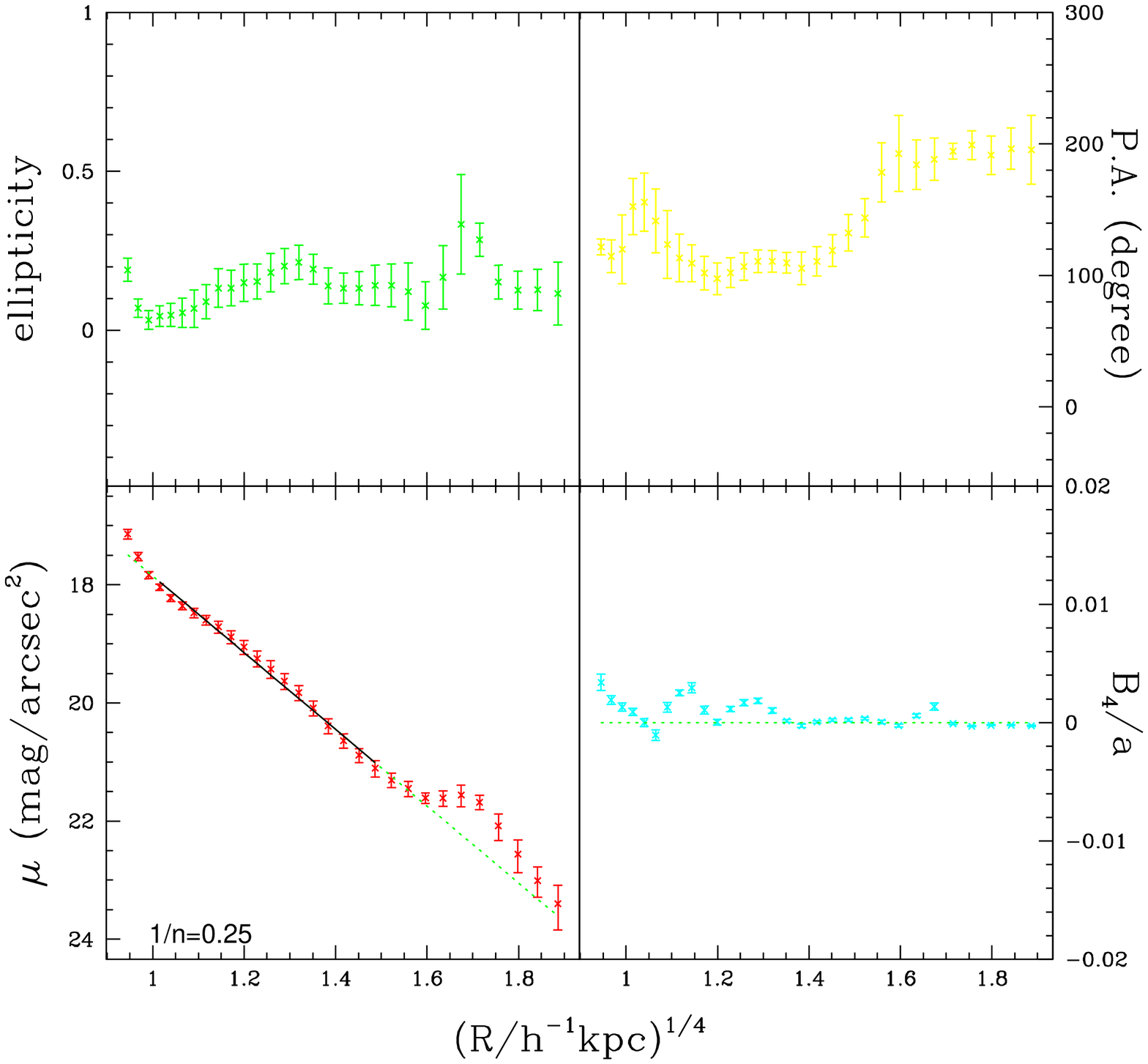}}
\caption{The left panel shows the HST Wide Field Camera image for
\Ba (the object labeled as G0).
The image shown is $74.5\arcsec$ on a side. A satellite
galaxy (G1) and a foreground star S1 are visible.
The middle panel shows the contours centered on \Ba; the contour 
levels are 1.7, 3.4, 6.8, 13.6, 27.2, 54.4, 108.8, 217.6, 435.2, 870.4,
1740.8 ADU, respectively. The thick
horizontal bar indicates a scale of 5$\hkpc$. The north and
east directions are indicated at the top right in the middle panel. The right
panels show the variations of the
ellipticity, position angle, 
surface brightness, and $B_4/a$ (see \S 3.3) as a function of \r14.
The best \r14 fit to the surface brightness profile
is given by the straight line.
}
\end{figure*}

\begin{figure*}
\resizebox{5.50cm}{!}{\includegraphics{url.ps}}
\resizebox{5.95cm}{!}{\includegraphics{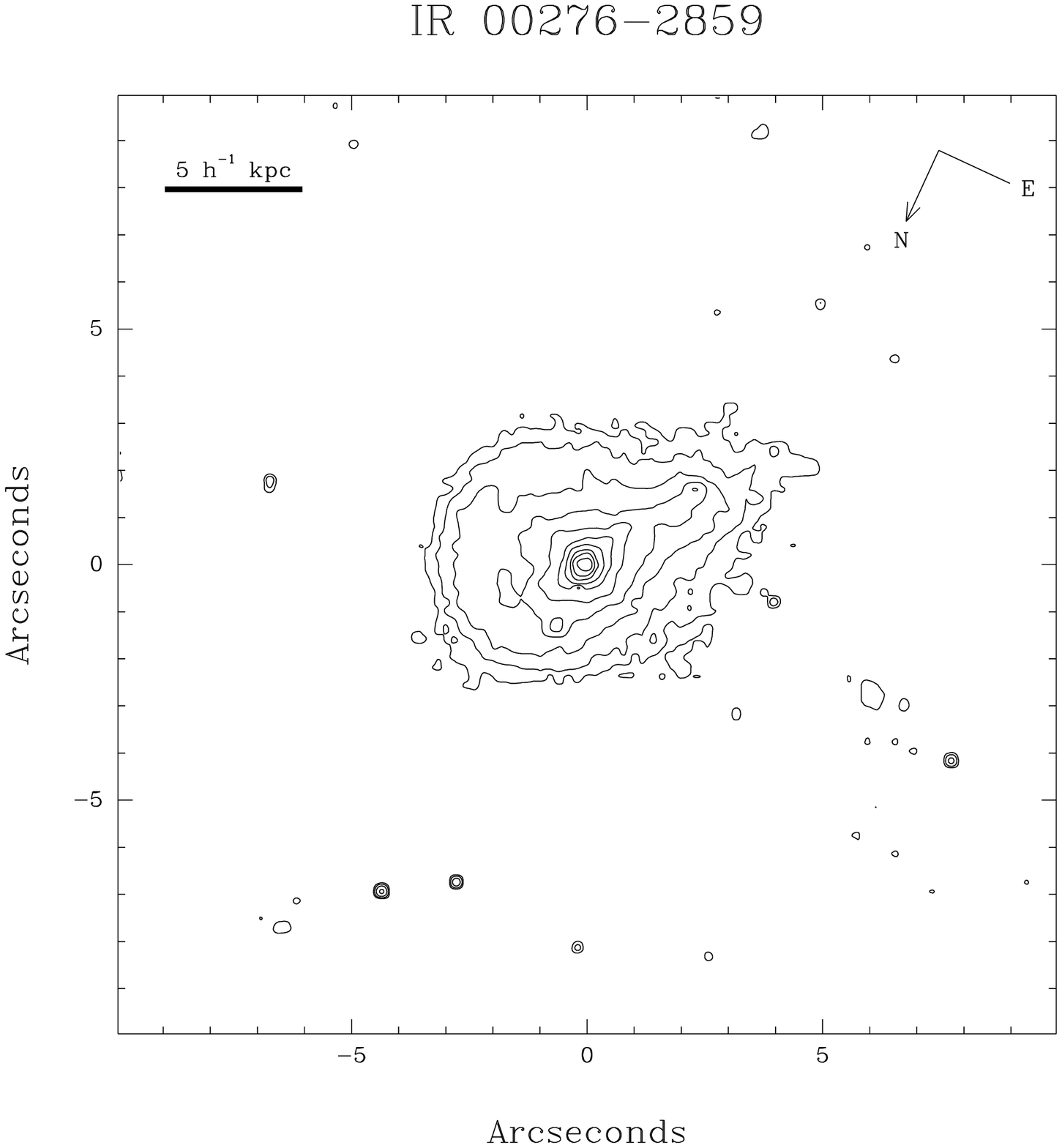}}
\resizebox{6.10cm}{!}{\includegraphics{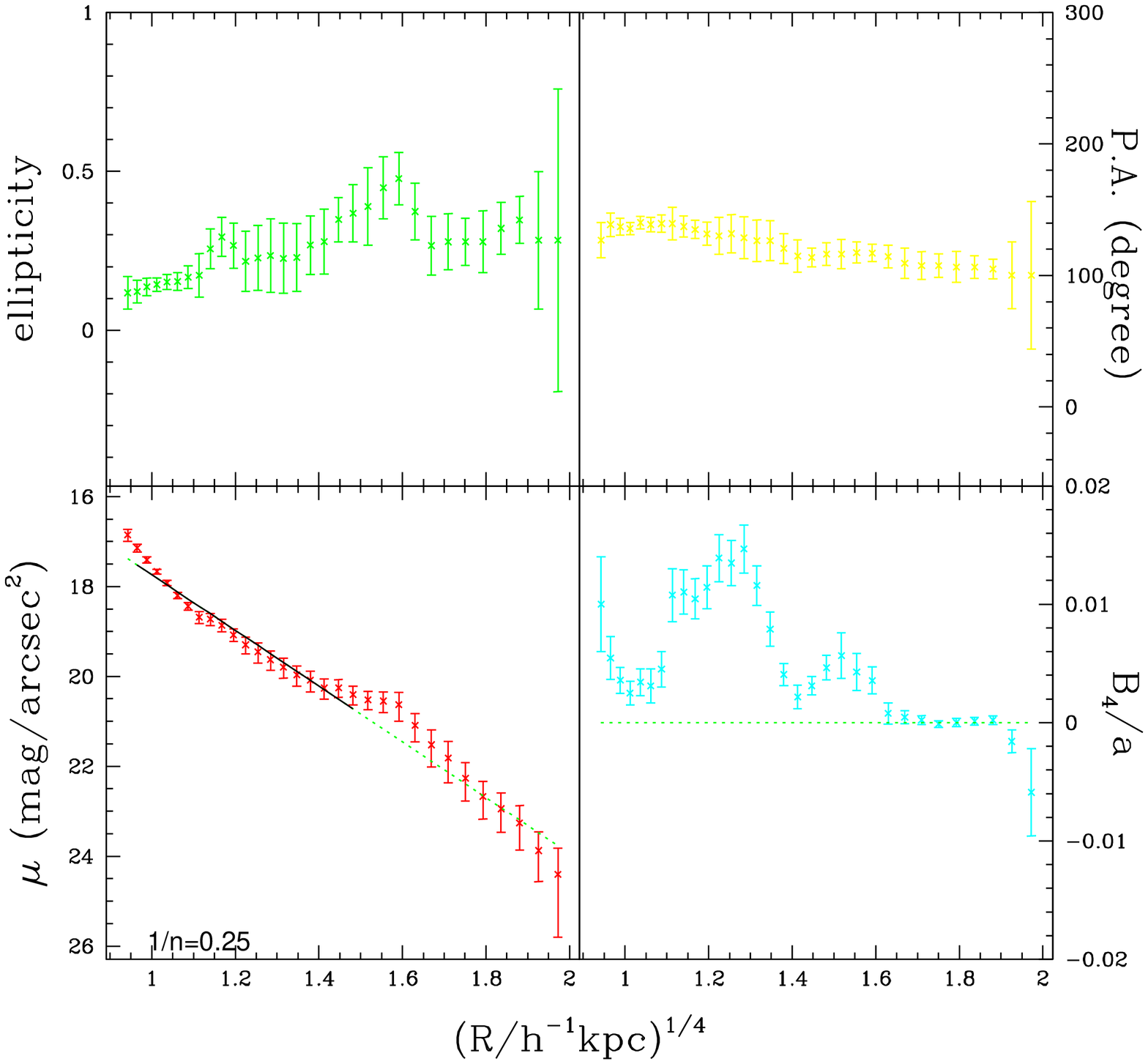}}
\caption{Same as in Fig. 5, for \Bb. A number of galaxies (G1-G5) and 
a star (S1) are within the field. The contour levels are
1.65, 3.3, 6.6, 13.2, 26.4, 52.8, 105.6, 311.2, 622.4, 1244.8, 
2489.6 ADU, respectively.
}
\end{figure*}

\begin{figure*}
\resizebox{5.50cm}{!}{\includegraphics{url.ps}}
\resizebox{5.95cm}{!}{\includegraphics{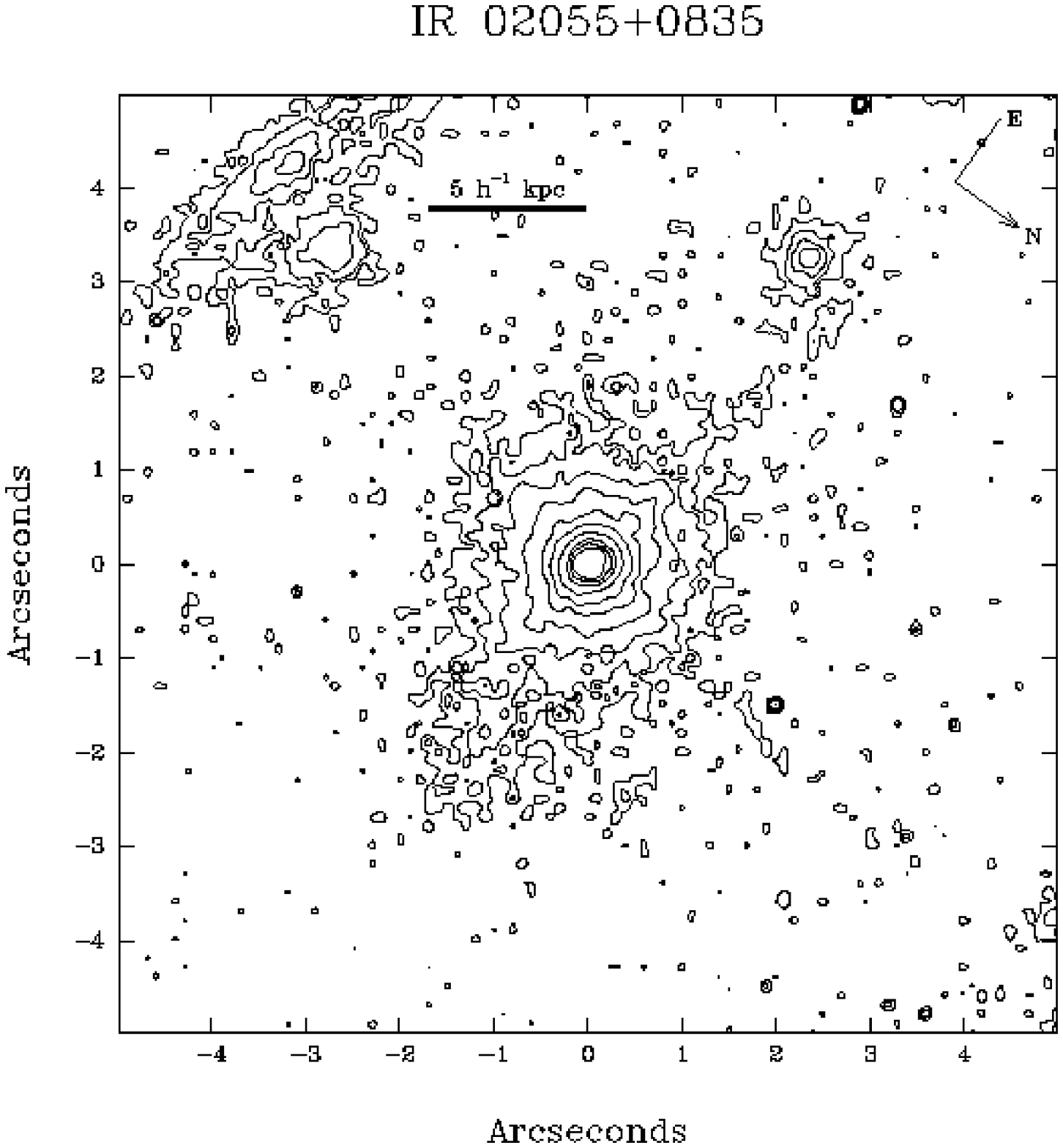}}
\resizebox{6.10cm}{!}{\includegraphics{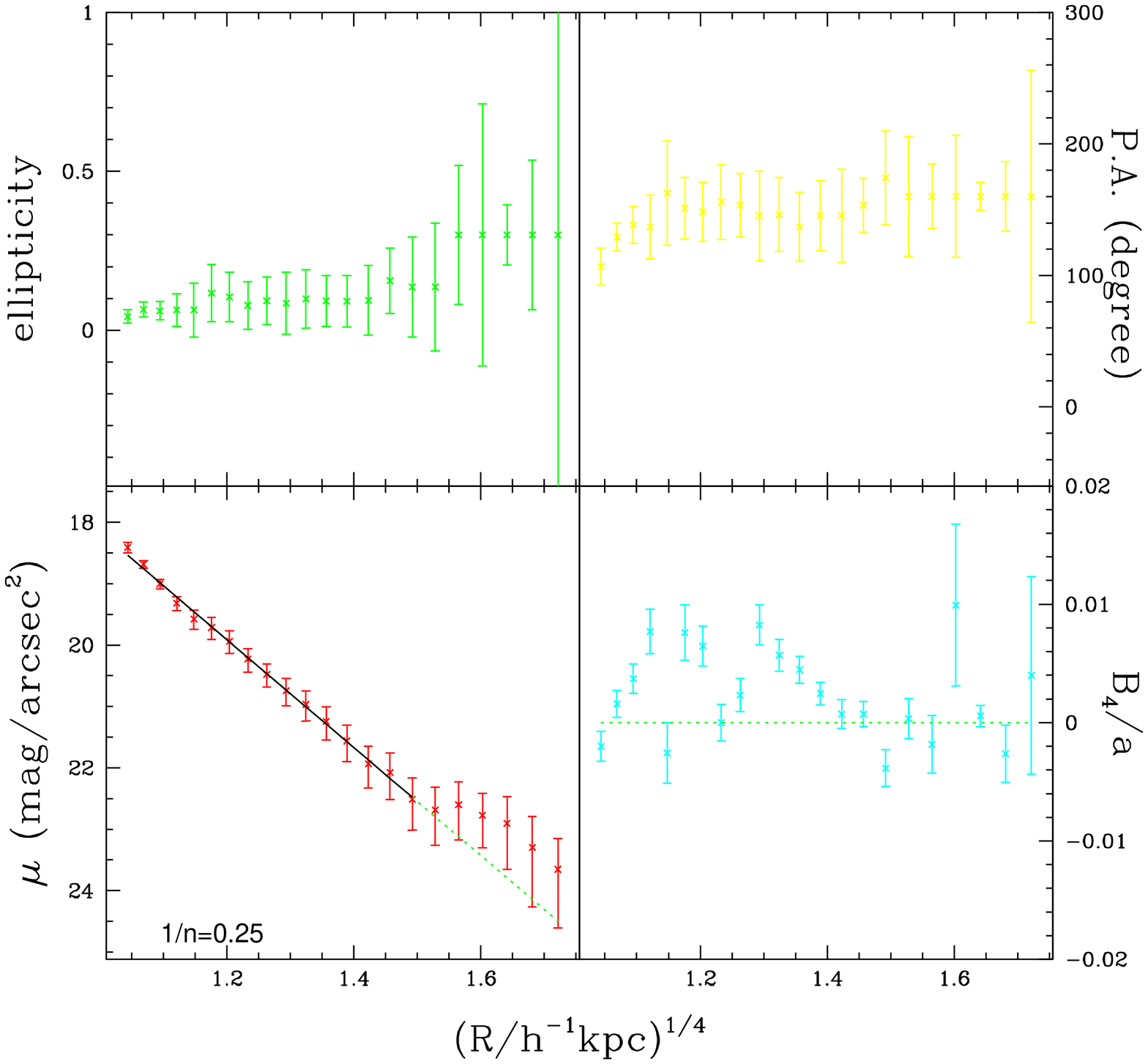}}
\caption{
Same as in Fig. 5, for \Bc.
The contour levels are
1.7, 3.4, 6.8, 13.6, 27.2, 54.4, 108.8, 217.6, 435.2, 870.4 ADU, respectively.
Two nearby objects (G1 and G2) are clearly visible.
}
\end{figure*}

\begin{figure*}
\resizebox{5.50cm}{!}{\includegraphics{url.ps}}
\resizebox{5.95cm}{!}{\includegraphics{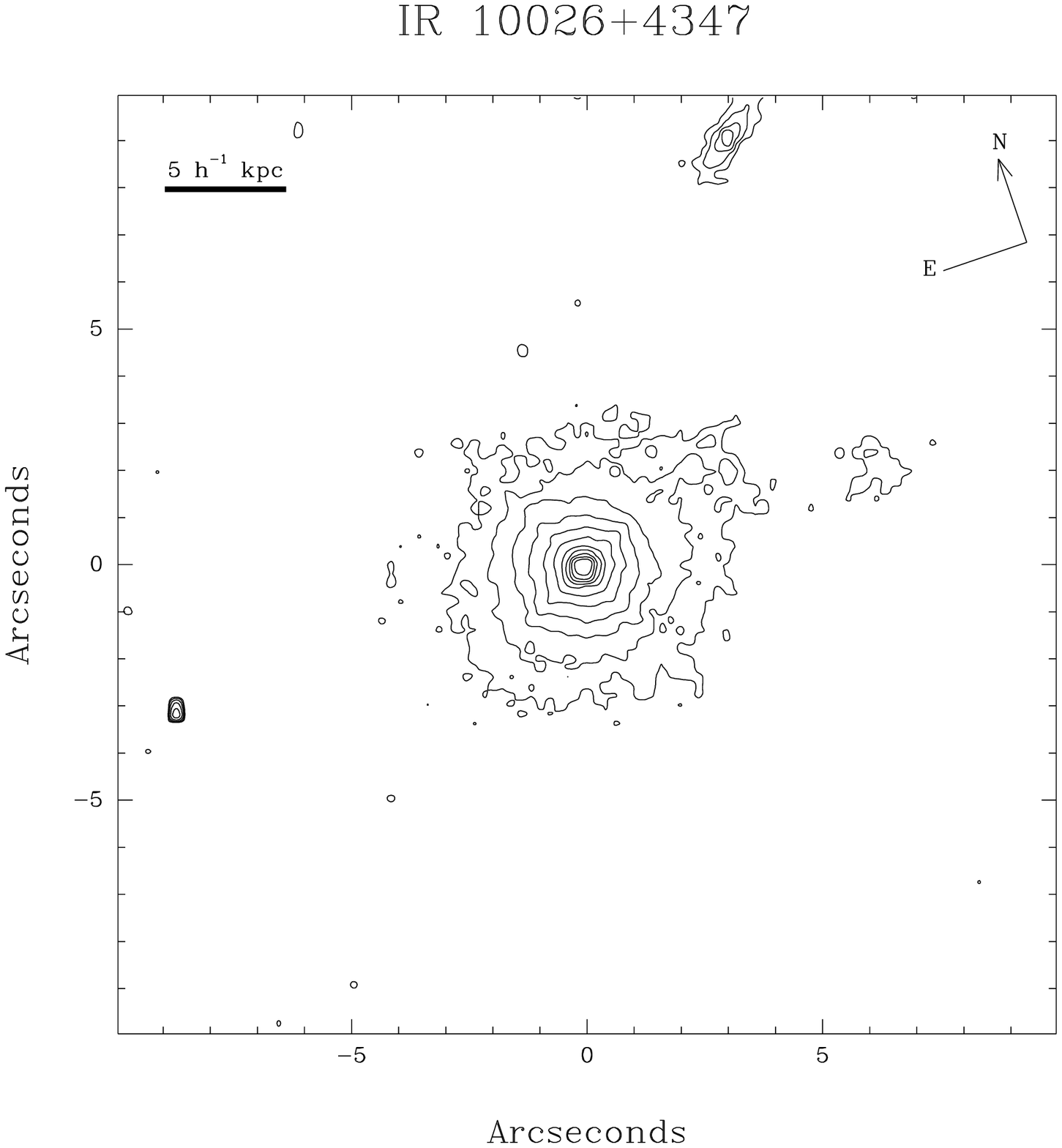}}
\resizebox{6.10cm}{!}{\includegraphics{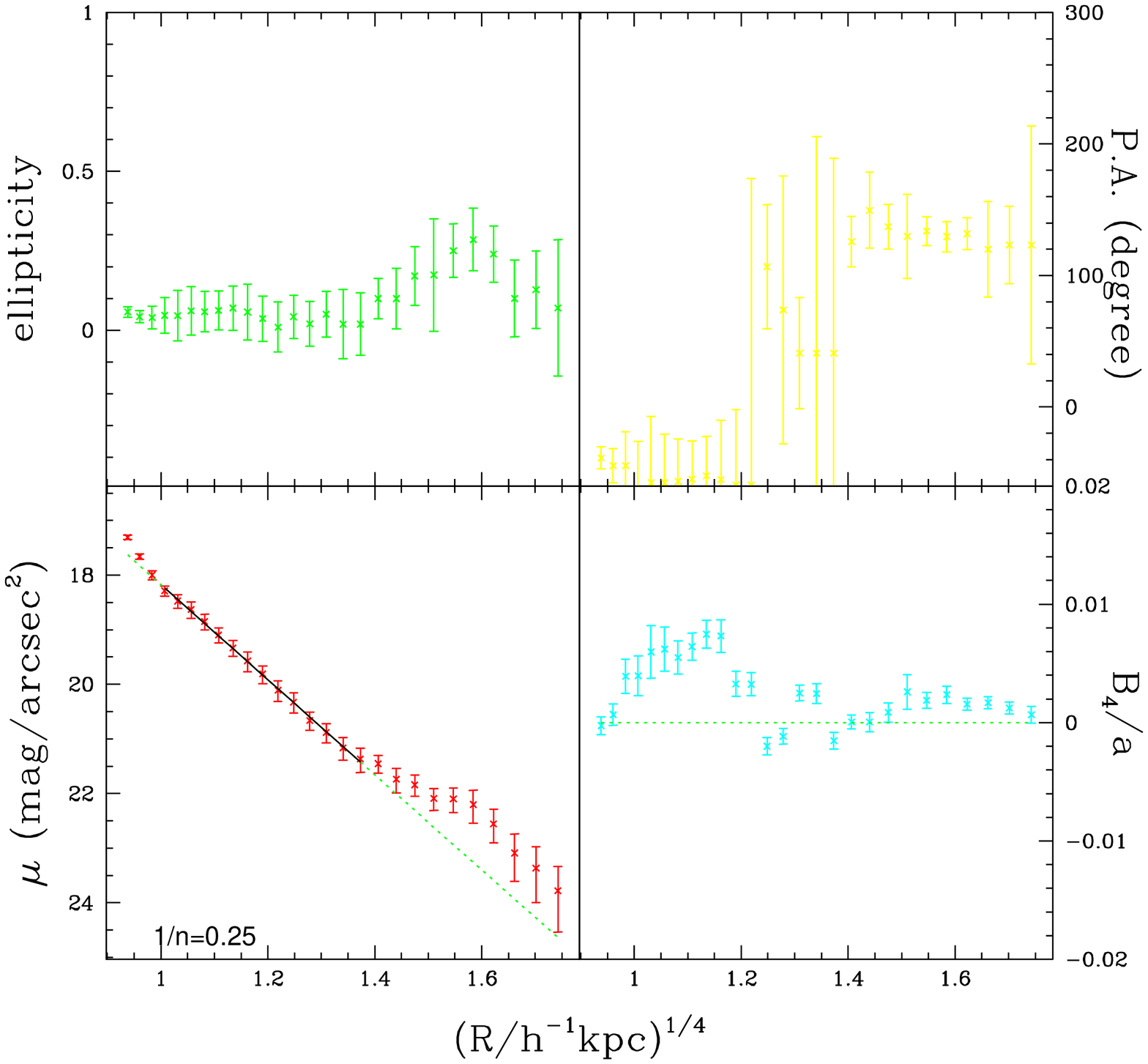}}
\caption{
Same as in Fig. 5, for \Bd. The contour levels are
1.75, 3.5, 7, 14, 28, 56, 112, 224, 448, 896, 1792 ADU, respectively.
}
\end{figure*}

The physical significance of the disky isophotes at the inner radius 
($\leq 1$$\hkpc$) and the excesses at ($\sim$ 3$\hkpc$) is not clear.
Multiple-waveband observations of ULIRGs reveal the presence of
massive starbursts in a molecular ring or disk
at several hundred pc to kpc in the circumnuclear regions
(e.g., Downes \& Solomon 1998).
The positions of the inner disky isophotes may reflect the presence of
such small disks at the center. Such nuclear disks are also found in
numerical simulations where gas is driven
inward to the central kpc region and cool efficiently at early stage
of galactic encounters (Icke 1985; Noguchi 1988). 
Therefore, the inner ``disky''
structure could be the relics of nucleus starburst rings or disks. 
The significance of bumps at $\sim$ 3$\hkpc$ is less clear, 
since even ``normal'' elliptical galaxies do not follow the \r14 law
perfectly (e.g., Saglia et al 1997). These departures could be produced
in the merging process by tidal force and/or dynamical friction.

\subsection{Class II}

Figs. 5-8 show the four galaxies in class II. Surface brightness profiles of
all these galaxies can be fitted by the \r14 law from
$\sim 1h^{-1}$ kpc out to $\sim 5h^{-1}$ kpc. The outer parts of these
galaxies show clear excess over the \r14 law.
We have tried a two-component fitting with an inner \r14 component
and an outer exponential component.
We find, however, that the fitting is unsatisfactory, meaning that
the outer extension is more complex than a simple exponential.
Disky isophotes are found in the inner region of each galaxy.
However, this is clearly an artefact caused by the saturated central
nuclei, since the diagonals of the quasi-diamond shaped isophotes
are almost along the diagonals of the images.

Although our classification of these galaxies are based purely on their
surface brightness profiles, remarkably, we found that all the four
galaxies host luminous bright nuclei and are saturated in the frames.
All the galaxies were
classified as Seyfert 1 galaxies spectroscopically (Lawrence et
al. 1999; Vader \& Simon 1987;
Moran et al 1996), and all have luminosities in the QSO regime. For
details, see Sect. 5.2.

\begin{description}
\item[\bf \Ba]
It has a small galaxy with a projected distance of $13\hkpc$ away to the west.
Two armlike features are clearly visible. The optical feature from
ground based deep observation looks like that of Mrk 231 (for details
see Surace et al 1998). These armlike features could be tidal
tails resulting from merging.

\item[\bf \Bb] There are a number of small galaxies within the same field.
In the contour plot, a small galaxy (G1) to the southeast can 
be detected. G1 causes
the outer contours of \Bb\, to protrude in its direction,
indicating that these two galaxies are physically interacting. 

\item[\bf \Bc] This galaxy  
has an extension to the west. Two objects G1 and G2 are within the same
field. Both objects are within a projected distance of 20$\hkpc$ and 
could be physically associated with the IRAS galaxy. Many
small and fainter objects are also visible within $20\arcsec$.

\item[\bf \Bd] This galaxy is an IR QSO and is potentially a member 
of one important class of transition
objects. Detailed discussions for this object
will be found in Xia et al (1999).

\end{description}

\begin{figure*}
\resizebox{5.50cm}{!}{\includegraphics{url.ps}}
\resizebox{5.95cm}{!}{\includegraphics{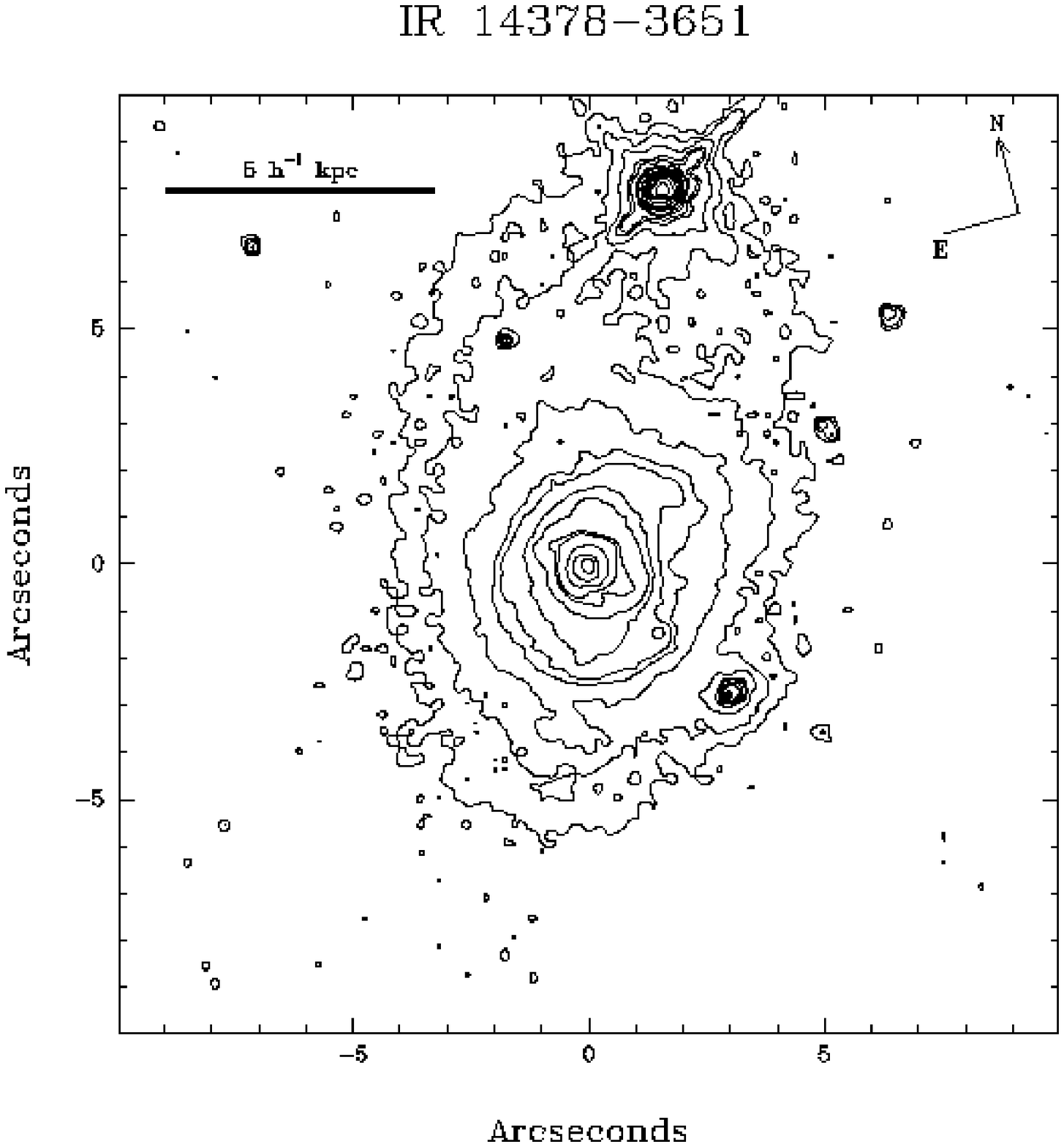}}
\resizebox{6.10cm}{!}{\includegraphics{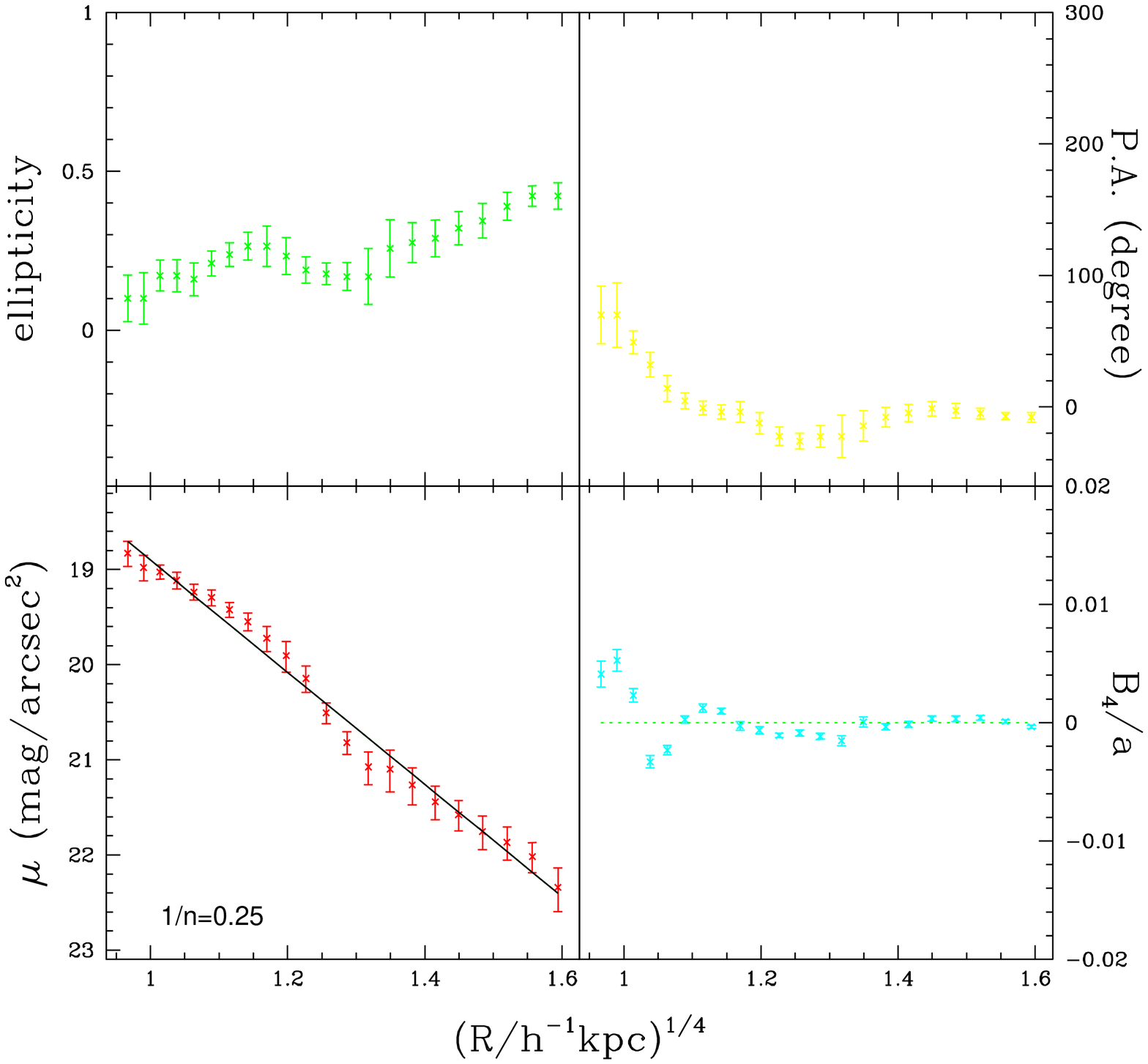}}
\caption{
The left panel shows the HST Wide Field Camera image for
\Ca (the object labeled as G0).
The image shown is $74.5\arcsec$ on a side. A satellite
galaxy (G1) and a foreground star S1 are visible.
The middle panel shows the contours centered on \Ca; the contour 
levels are 
1.9, 3.8, 7.6, 15.2, 20, 30.4, 40, 60.8, 70, 121.6, 243.2, 486.4, 
972.8, 1943.6 ADU, respectively. The thick
horizontal bar indicates a scale of 5$\hkpc$. The north and
east directions are indicated at the top right in the middle panel. The right
panels show the variations of the
ellipticity, position angle, 
surface brightness, and $B_4/a$ (see \S 3.3) as a function of \r14.
}
\end{figure*}

\begin{figure*}
\resizebox{5.50cm}{!}{\includegraphics{url.ps}}
\resizebox{5.95cm}{!}{\includegraphics{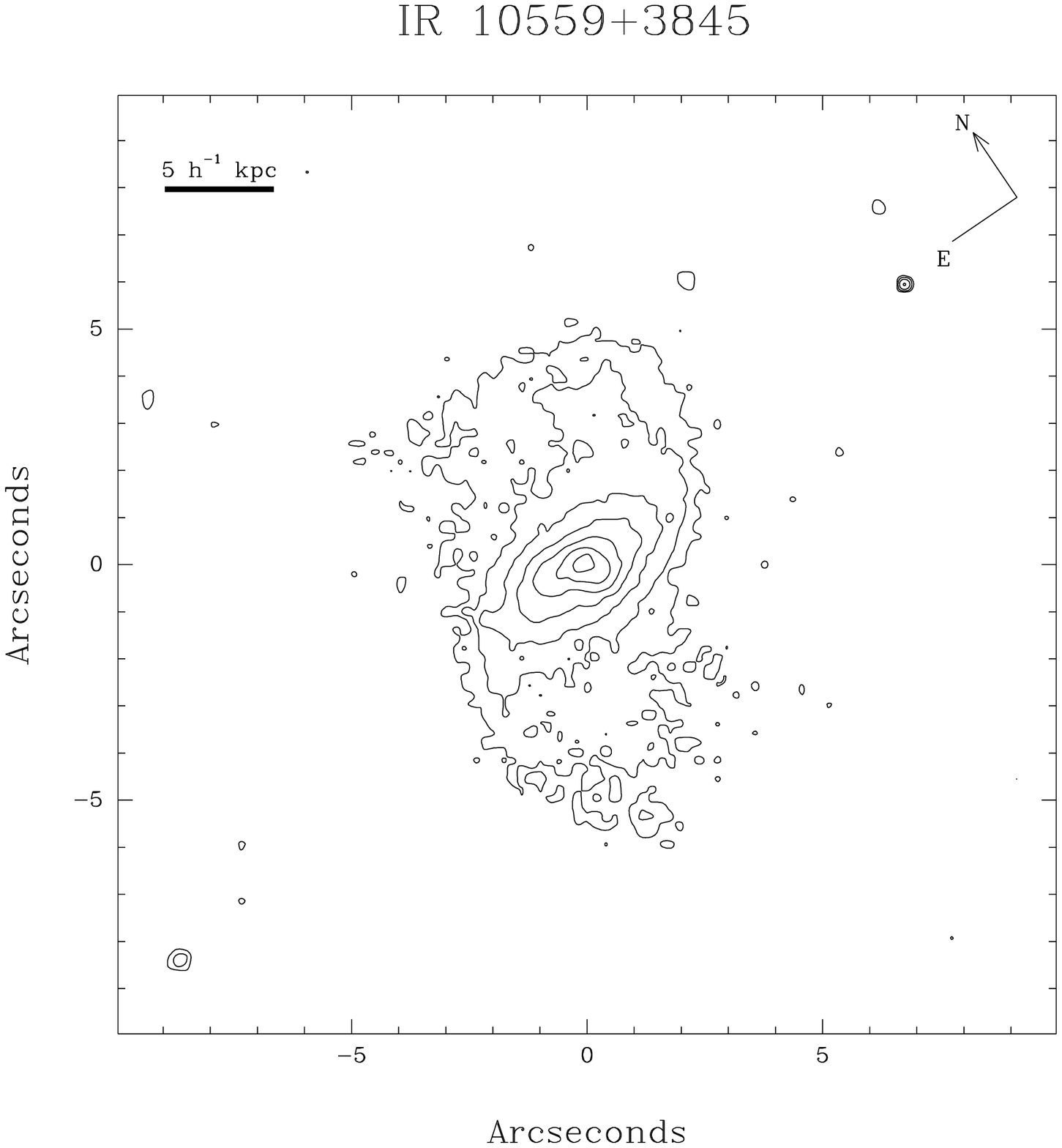}}
\resizebox{6.10cm}{!}{\includegraphics{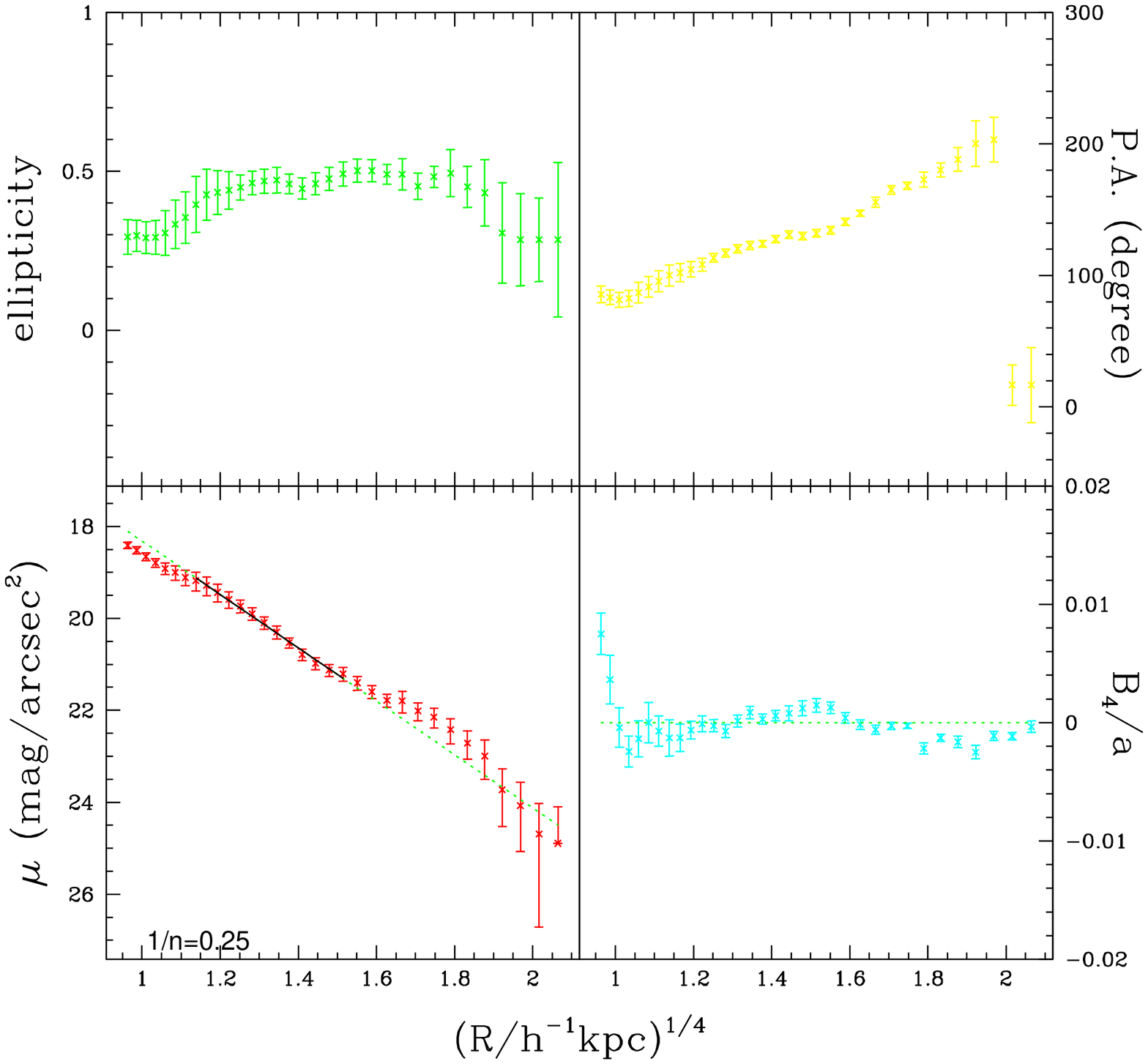}}
\caption{
Same as in Fig. 9, for \Cb. The contour levels are 
1.7, 3.4, 6.8, 13.6, 27.2, 54.4, 108.8, 217.6, 435.2 ADU, respectively.
}
\end{figure*}

\begin{figure*}
\resizebox{5.50cm}{!}{\includegraphics{url.ps}}
\resizebox{5.95cm}{!}{\includegraphics{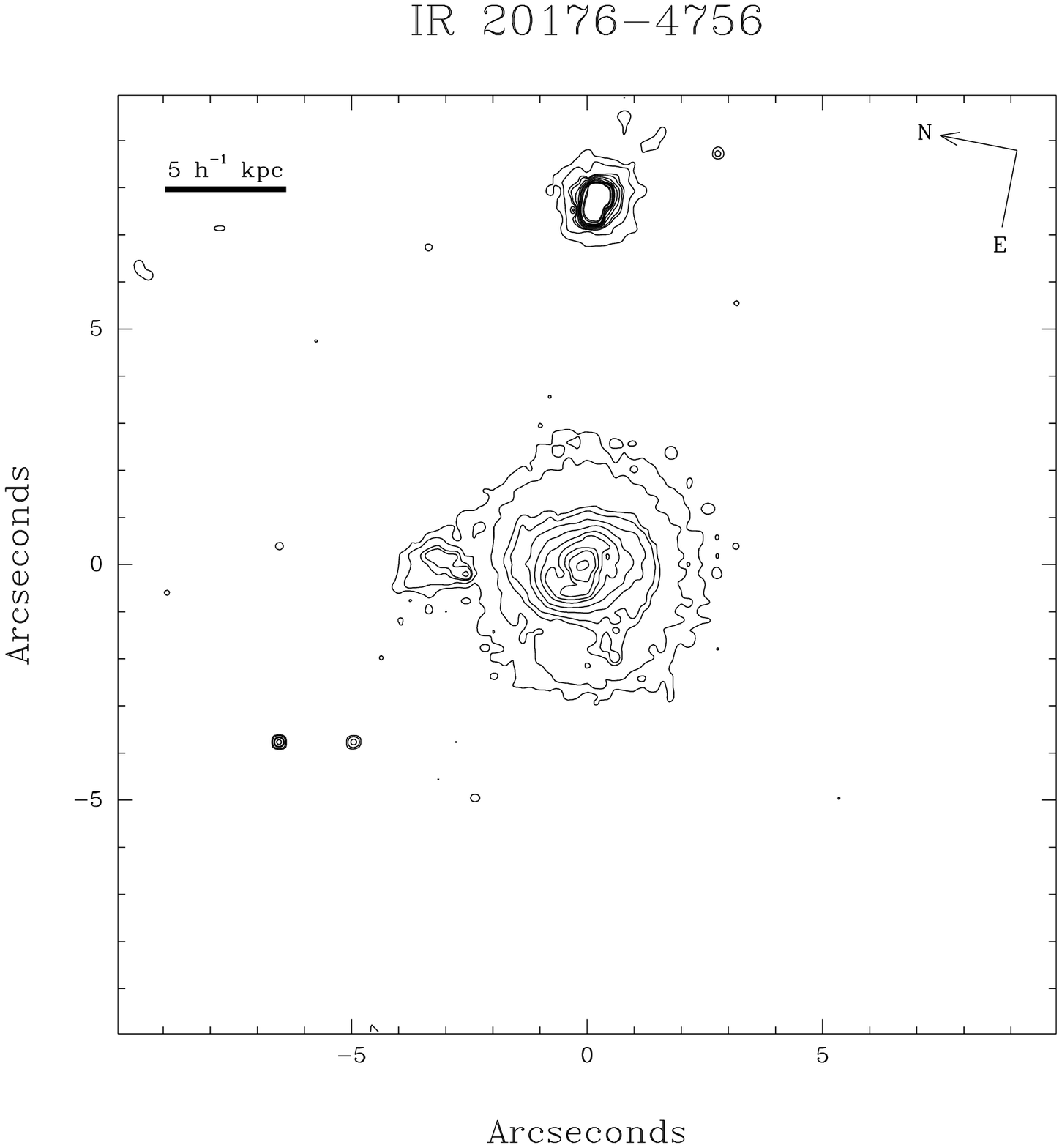}}
\resizebox{6.10cm}{!}{\includegraphics{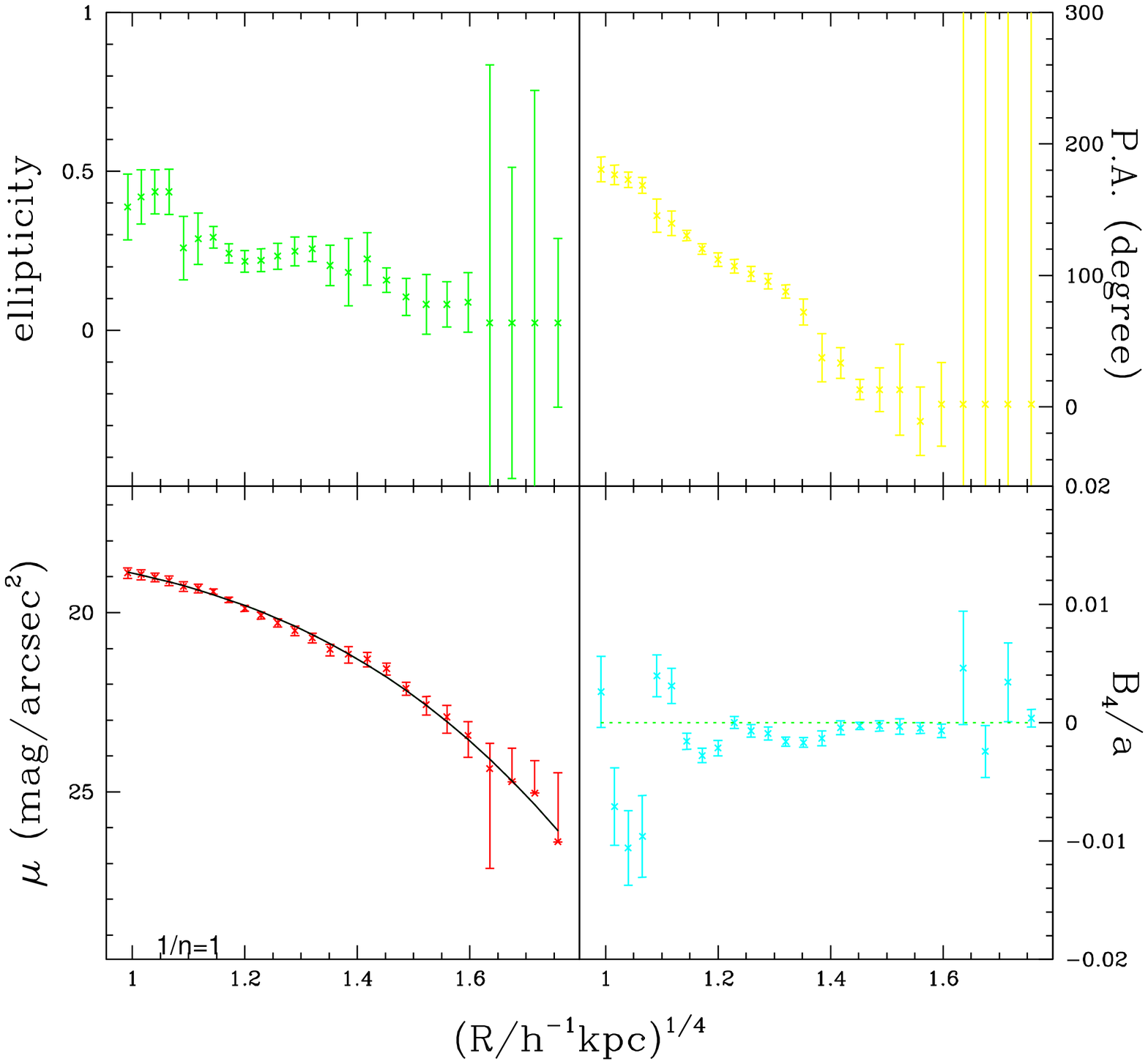}}
\caption{
Same as in Fig. 9, for \Ce. The contour levels are 
1.75, 3.5, 7, 9, 14, 20, 28, 35, 45, 56, 90, 112 ADU, respectively.
An excellent exponential fit to the surface brightness profile is 
also plotted. The arrow marks the protrusion and
and thick tail-like material in the outer region.
}
\end{figure*}

\begin{figure*}
\resizebox{5.50cm}{!}{\includegraphics{url.ps}}
\resizebox{5.95cm}{!}{\includegraphics{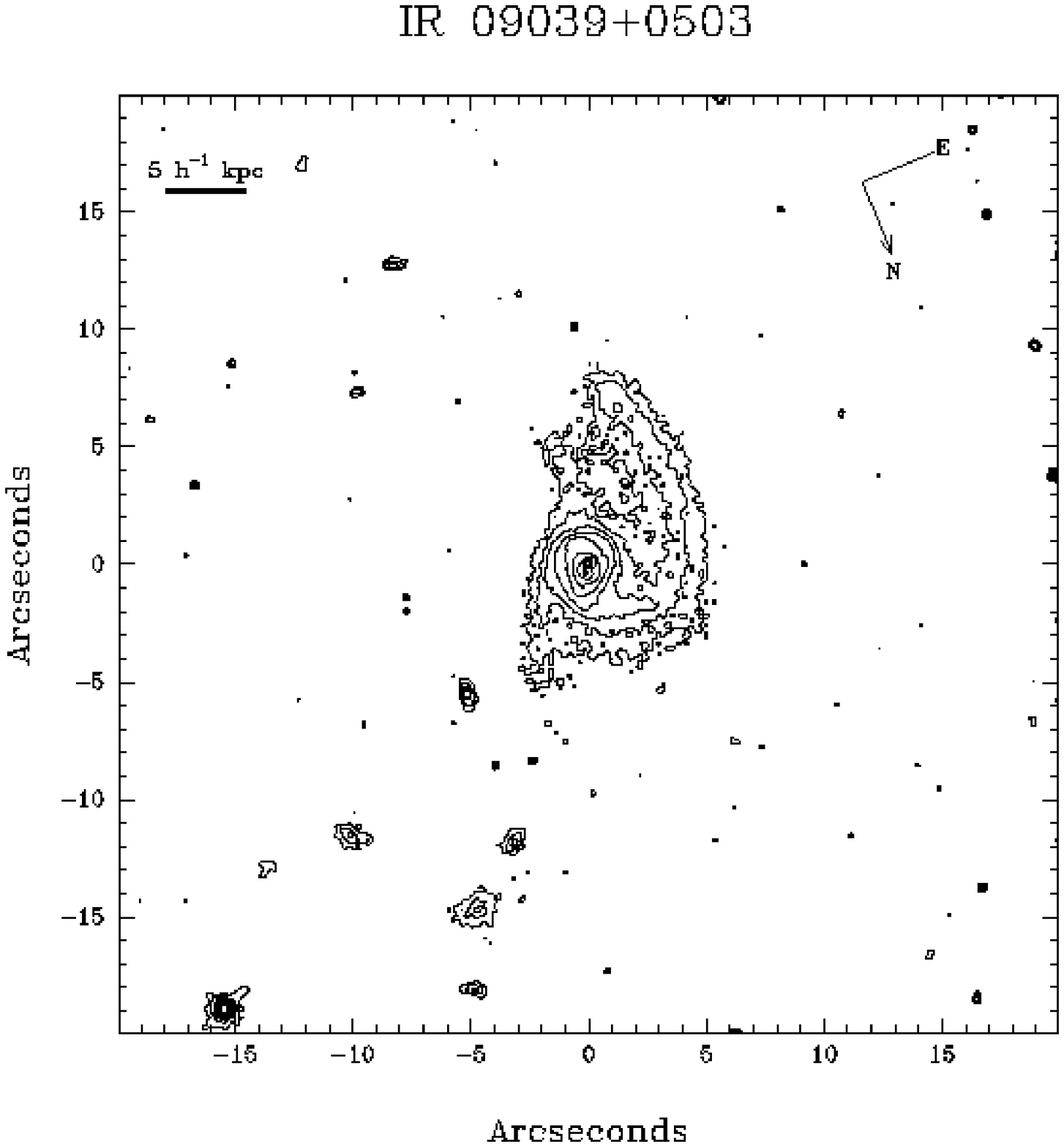}}
\resizebox{6.10cm}{!}{\includegraphics{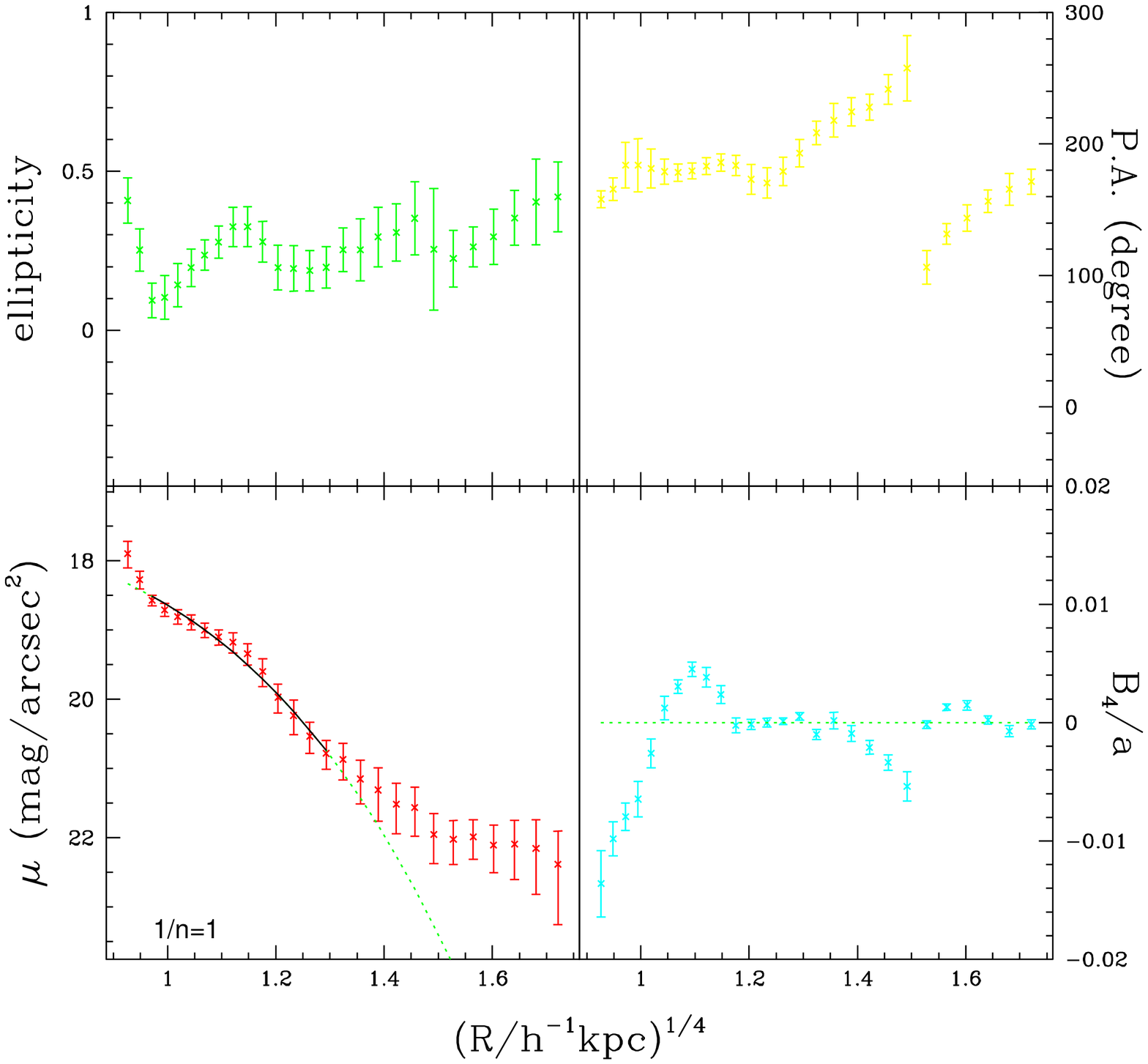}}
\caption{
Same as in Fig. 9, for \Cc. The contour levels are
1.77, 3.54, 7.08, 14.16, 28.32, 56.64, 113.28, 200, 226.56 ADU, respectively.
An exponential  fit to the surface brightness profile is also plotted.
}
\end{figure*}

\begin{figure*}
\resizebox{5.50cm}{!}{\includegraphics{url.ps}}
\resizebox{5.95cm}{!}{\includegraphics{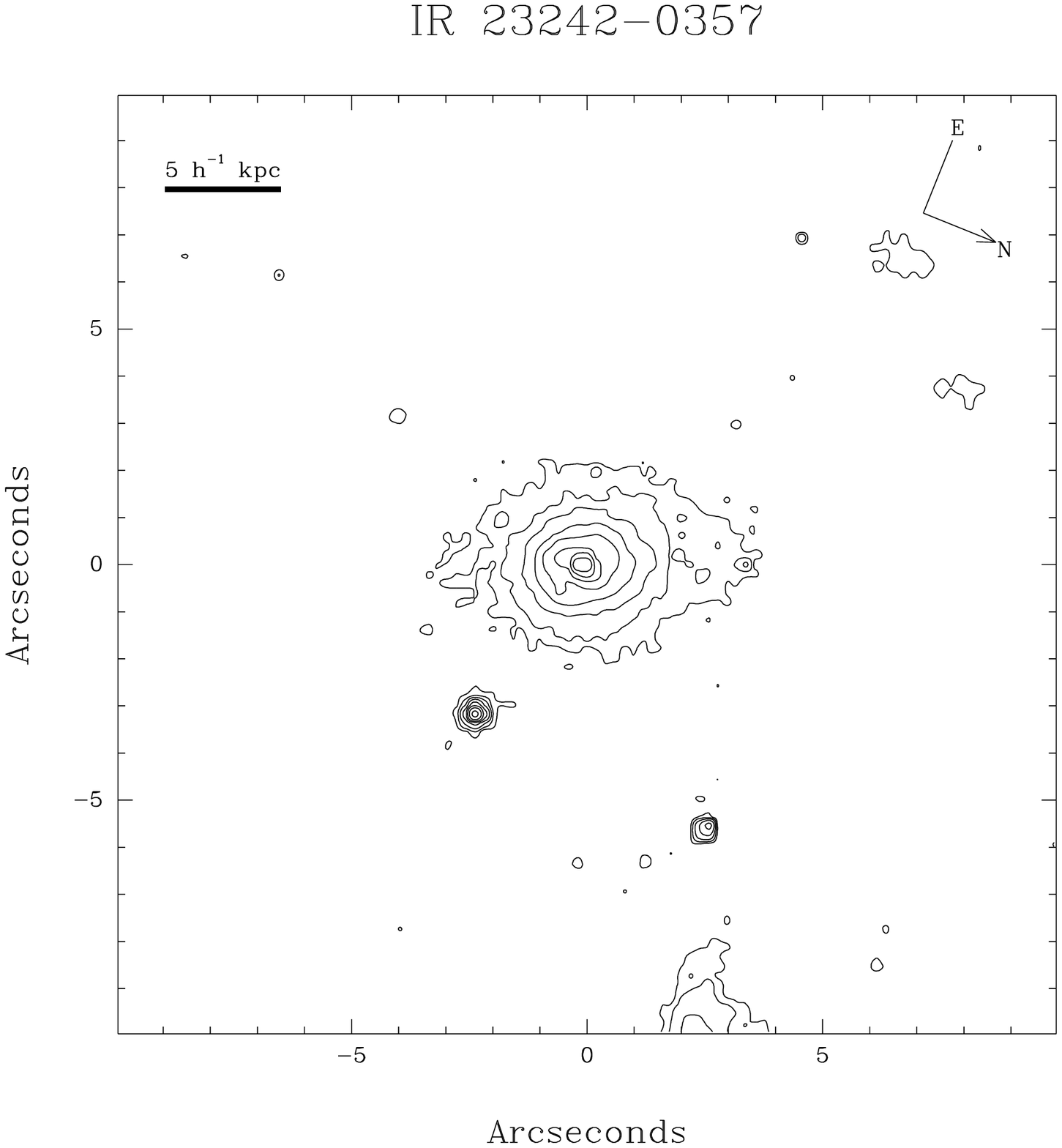}}
\resizebox{6.10cm}{!}{\includegraphics{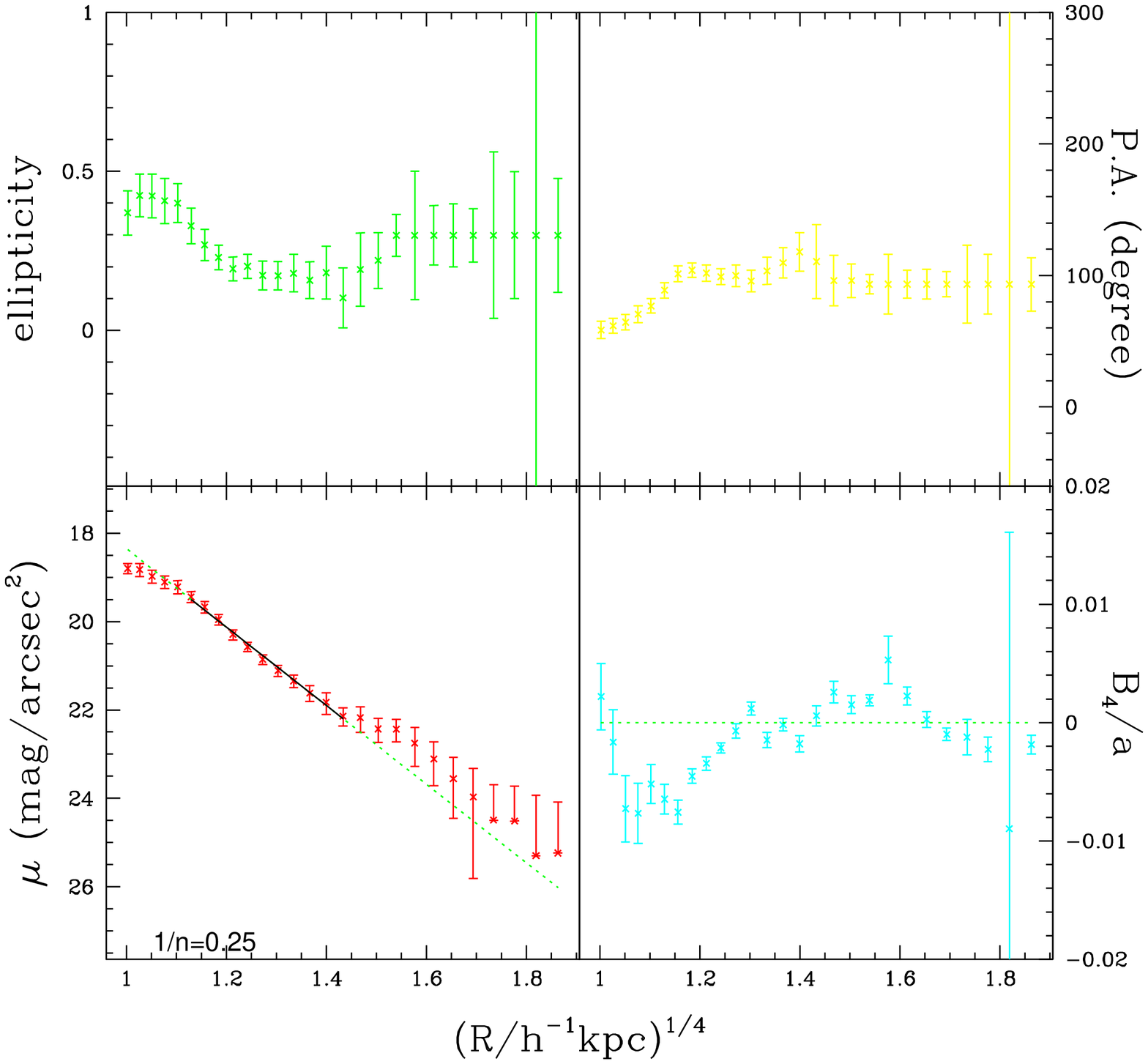}}
\caption{
Same as in Fig. 9, for \Cd. The contour levels are
1.7, 3.4, 6.8, 13.6, 27.2, 54.4, 108.8, 217.6, 435.2, 870.4 ADU, respectively.
There are four other galaxies (G1-G4) within the field. The best \r14 fit 
to the surface brightness profile
is given by the straight line.
}
\end{figure*}

\subsection{Class III}

The five galaxies in this class have different surface 
brightness profiles, and so we discuss them in turn. 

\begin{description}
\item[\bf \Ca]
It is a galaxy with an asymmetric extension, being more diffuse to the 
northwest, as can be seen in Fig. 9. A small galaxy (G1) is visible 
about 4$\hkpc$ (projected) to the southwest.
This galaxy is still within the 
envelope of \Ca and could be physically associated with
\Ca. A ripple-like material can be separated from 
the southeast part of the extension.
A loop structure can be seen in the 
inner part which is similar to spiral arms. The isophotes 
are disky within 1$\hkpc$. The surface brightness profile
of this source deviates from 
the \r14 law. If we fit it with the \r14\ law, we find a bump at 
$1.5h^{-1}$ kpc and a dent at 
$2.8h^{-1}$ kpc. The inner bump is related to the inner arms.
Apart from the outer part, 
this galaxy has a round appearance ($\epsilon < 0.3$). The ripples,
loop structures and the deviation from the \r14 law suggest that
that \Ca\ is a merging galaxy and is still relaxing towards 
the \r14 light distribution.

\item[\bf \Cb]
  From Fig. 10, \Cb\ appears to have
two nearly symmetric diffuse tails. NICMOS observations reveal
two very close nuclei (Borne et al 1998), indicating that this galaxy is a
merging galaxy as well.
The position angle of the major axis continuously changes by
about 120 degrees from the center to the outer part; 
the position angle change in the inner part is more gradual
than that in the outer part.
The ellipticity of the inner isophotes is smaller (about 0.3) 
than the outer value (about 0.5). 
The center of the inner isophotes offsets to the south with respect
to that of the outer ones. 
The mergering process appears not yet complete for this galaxy.

\item[\bf \Ce]
  From both image and surface brightness profile (see Fig. 11),
this galaxy appears to have been well relaxed. Remarkably,
the surface brightness profile can be well fitted by a 
single exponential, ranging from $0.8\hkpc$ to $9.5\hkpc$.
The central part ($\leq 1.4 \hkpc$) has boxy isophotes.
On the other hand, there still exist merger signatures, such as
a loop structure (or inner arms) in the inner region, fan-like protrusion
and thick tail-like material at the outer region (marked with an arrow in
the image). The most significant isophotal feature of \Ce\ is 
that the position angle gradually decreases by about $180^\circ$. 
We speculate that the central boxy isophotes and isophotal twists
may be caused by a central bar.

\item[\bf \Cc] This galaxy
has a remarkable morphology as shown in Fig. 12.
A thick and luminous tail emerges
from the north
and bends to the south with a total length of about $\sim 15\hkpc$.
Two short tails locate to the south and northwest, respectively.
A closer look reveals that this galaxy has two very close nuclei, 
which gives rise to the boxy isophotes within 1.2$\hkpc$.
Although the central parts of the two galaxies 
have almost merged together, the galaxy is still far from  
complete relaxation and the surface brightness 
profile deviates significantly from the \r14\ law.
There are many small and faint objects within 
$30\arcsec$ of the source. It would be 
interesting to determine the redshifts of these small galaxies to see whether
they are physically associated with the source.

\item[\bf \Cd]
It has a surface brightness profile similar to that of the second class, it
has an inner 
component satisfying \r14\ law and a more extended outer component. The
primary difference is that this galaxy has
a fainter nucleus. The inner region has two fan structures 
which cause the boxy isophotes at radius $\leq 2.8\hkpc$). 
The galaxy protrudes 
to the northeast and has a faint and slim spiral tail around it in the
south. There are also four fainter galaxies (G1-G4) within the field, 
indicative of a group environment. Many fainter objects can be found
within $15\arcsec$ of the source.

\end{description}

To summarize, the five galaxies
in class III have diverse appearances,  including one (\Ca) with a close
double nuclei. Their surface brightness
profiles deviate to different degrees from the \r14 law, with \Ce\,
fitted well by an exponential.
\Cd\, is similar to the objects in class II except that
it has a fainter nucleus.
We will return to the issue of the connection between the three classes
of objects in the last section.

\section{The Optical Spectroscopic Properties}

Our studies so far have concentrated on the photometric properties
of our sample galaxies. Clearly it is important to explore the spectroscopic
properties of these galaxies in order to further understand
these three classes of galaxies. In order to address 
this question and also to understand the possible group environment
which we alluded to in Section 4, we have carried out spectroscopic
observations from Dec. 20 to Dec. 23, 1998 using
the 2.16m telescope at the Xinglong Station of Beijing Astronomical
Observatory.
We have observed \Ac, \Aa, \Cc\, and \Cd, using
a Zeiss universal spectrograph mounted on the 2.16m telescope.
A Tektronix 1024$\times$1024
CCD was used giving a wavelength coverage of 
4100A to 9100A
with a grating of 200A/mm. The spectral resolutions are 9.3A (2 pixels).
Wavelength calibration was carried out using a He-Ar lamp;
the resulting wavelength accuracy is better than 1A.
KPNO standard stars were observed to perform flux calibrations.

 In the following section, we will present our observational
results together with data available in the literature, such as 
Kim et al (1998) and  Lawrence et al (1999), for these
three classes of objects.

\subsection{Class I}

\begin{figure}
\resizebox{\hsize}{!}{\includegraphics{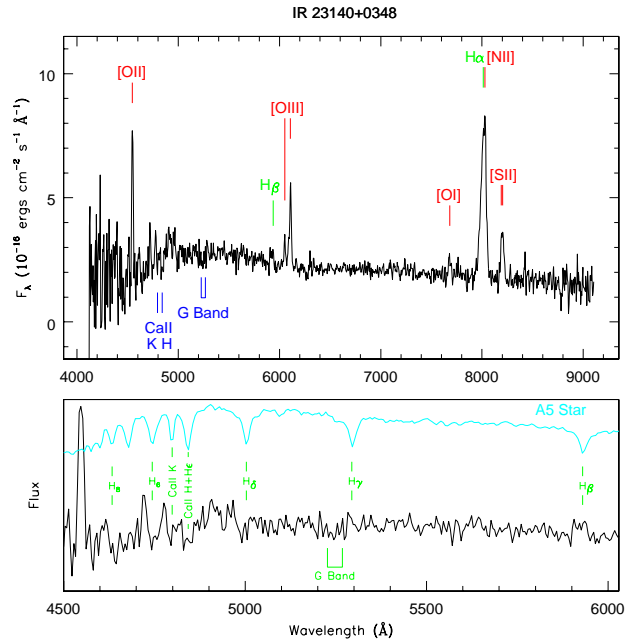}}
\caption{
The top panel shows the optical spectrum for \Ac. The bottom panel shows
a magnified view for the spectral region between 4500A to
6000A. For comparison of absorption lines,
the spectrum of an A5 star is shown. Prominent lines are labelled.
}
\end{figure}

The spectrum of \Ac\, is shown in Fig. 14. As can be seen from this
figure, there are strong emission lines, as well as
obvious 4000A break (redshifted to $\sim$4900A),
G-band and possible CaII K and H absorption features.
The Balmer absorptions are marginal in this low signal-to-noise
ratio spectrum. From the emission line ratios and the classification criteria
(Veilleux et al 1995; Osterbrock et al 1989), \Ac\ can be classified as
a LINER. 

Compared with the ``E+A''-like spectrum of ULIRGs FF J1614+3234 and TF 
J1020+6436 (Tran et al 1999; Breugel 1999; see also Dressler \& Gunn 1983),
the Balmer absorption line features of \Ac\ is weaker.
Nevertheless, the spectral features resemble each 
other and it indicates that \Ac\ has a population of young stars with 
age $\sim 1$ Gyr. Combining with the fact that its surface brightness
profile is well described by the \r14 law, we conclude that
\Ac\ is a young elliptical
galaxy resulted from merging. This is further supported by the fact that
\Ac\ is a radio galaxy, and such galaxies 
are often associated with merger remnants (Barnes 1998).

We have also observed three galaxies
(G0, G1 and G2 in Fig. 1) in the field of \Aa.
We found that the redshifts for G0 and G2 are 0.284 and 0.149,
respectively; the low signal-to-noise ratio and the weak emission features
make it difficult to securely identify G1's redshift.
The redshift of G2 is the same as the redshift given for
\Aa\, by NED\footnote{
The NASA/IPAC Extragalactic Database (NED) is operated by
the Jet Propulsion Laboratory, California Institute of Technology,
under contract with the National Aeronautics and Space Admimistration.}.
The IRAS error ellipse (see Fig. 1) touches 
both G0 and G2, with G0 closer to the center of the error ellipse, so the
identification is somewhat ambiguous.
The spectrum, shown in Fig. 15, indicates that G0 is a narrow line
Seyfert 1 galaxy with extremely strong FeII emission.
Since almost all extremely strong
FeII emitters are ULIRGs, G0 (at $z=0.284$)
is most probably the optical counterpart of the ULIRG. The NED
identification seems to have confused the objects G0 and G2. 
Incidently G2 is a foreground galaxy at $z=0.149$ with an HII-region
like spectrum.

For the other two objects in class I, \Ad\ is also a Seyfert 1
galaxy with FWHM of \Halpha\, less than $3000 \kms$ (Lawrence et al 1999) 
while the spectral classification for \Ab\ is LINER. 

\subsection{Class II}

\begin{figure}
\resizebox{\hsize}{!}{\includegraphics{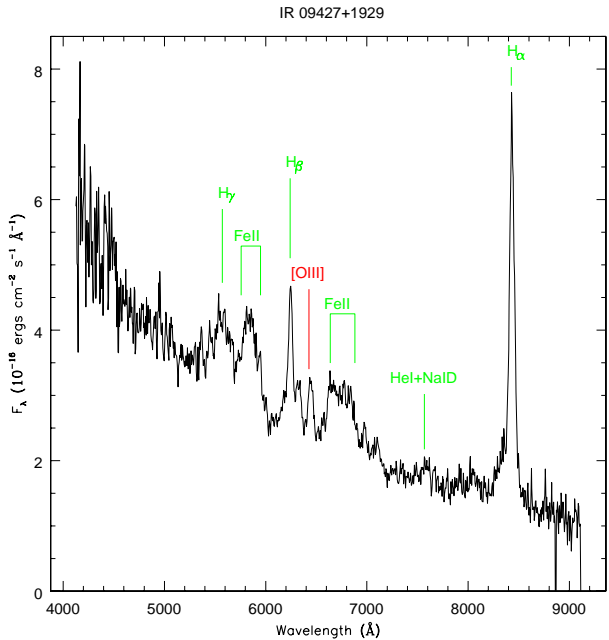}}
\caption{
Spectrum for \Aa. Prominent lines are indicated. Notice the strong FeII lines.
}
\end{figure}

As for the spectra of galaxies in 
Class II, \Ba\ (I Zw 1) is a typical narrow line
Seyfert 1 (NLS1) galaxy (Moran et al 1996); \Bd\ is a NLS1-like QSO
(Xia et al 1999); \Bb\ is an IR QSO with strong FeII emission line
and with FWHM of \Hbeta\ about $2000 \kms$; \Bc\ is also a QSO with
strong FeII emission line and the FWHM of \Hbeta\ is less than $3000 \kms$
from the QDOT redshit survey spectrum obtained by the William Hershel
Telescope (Lawrence et al 1999).
It is remarkable that all four galaxies in Class II
are classified as Seyfert 1 galaxies with strong FeII emission lines
and relatively narrow permitted emission lines. Furthermore, the optical
luminosities for all four galaxies are in the range of QSOs. 
Except for \Bc, all the other three objects were detected in the soft
X-ray by ROSAT (the non-detection of \Bc\, is perhaps because it is the
most distant galaxy in the sample). All the three soft X-ray spectra
are steep. Therefore, these four objects are NLS1-like QSOs 
(Boller et al 1996), or they are en route to normal QSOs
(Xia et al 1999).

\subsection{Class III}

The spectrum for \Cd\, is a typical HII region spectrum (see Fig. 16).
Table 2 gives the spectral classifications for the other 4 galaxies 
in this class based on the data from the
literature or the spectra from the QDOT redshift survey.
\Ca\ and \Cc\ are classified as LINERs based on Kim et al (1998) 
and our spectrum (not shown),
respectively. As Veilleux et al (1995) pointed out that the excitation
mechanism for infrared luminous LINERs could be shock waves. Therefore
it is possible that strong shocks are present
in \Ca\ and \Cc, as in some starburst galaxies,
such as NGC 6240 and NGC 3690 (Heckman et al 1990). 
To summarize, the statistical spectral properties of class III objects
are very different from those of Class I and Class II. These objects
have either HII region-like or LINER-like spectrum, consistent with
the presence of starbursts in 
these galaxies without obvious central AGNs. 

\begin{figure}
\resizebox{\hsize}{!}{\includegraphics{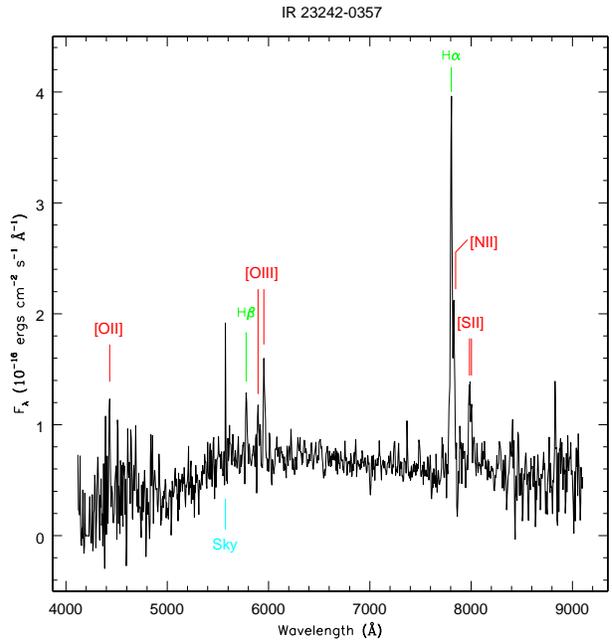}}
\caption{
Spectrum for \Cd. Prominent lines are labelled.
}
\end{figure}

\section{Summary and Discussions}

   From a total of 58 HST WFPC2 I-band snapshot images of ULIRGs, we have
selected 13 sources that are
dominated by a single nucleus to study their surface photometry.
According to their surface brightness profiles, we find that these
galaxies can be divided into three classes. Below we first
briefly summarize their
morphology, surface brightness profiles and optical properties, and then
discuss the origin and evolution tracks of these galaxies.

  The surface brightness profiles of the four galaxies in class I
are well fitted by the \r14\ law out to $\sim 10 \hkpc$. 
with bumps at inner few kpc regions.
The disky isophotes at the inner region $\la 1\hkpc$
are consistent with stellar disk formed by infalling gas 
during merging; molecular disks of comparable size 
have been detected in CO (e.g., Downes \& Solomon 1998).
The fine structures at the outer part of these galaxies hints
that the relaxation process is incomplete and
is slower than the inner part. 
The large position angle changes also hint
that these galaxies are merger remnants and are still 
evolving into ellipticals. Each member of this class has round appearance.
The round appearance suggests that they have experienced
multiple mergers, since binary merging tends to produce more elongated
objects (Weil \& Hernquist 1996). These objects
are at the last stage of merging. 

  The four galaxies in Class II are composed of an inner component 
described by the \r14\ law out to $\sim 5\hkpc$ plus some outer extension. 
The existing outer extension indicates that this class of ULIRGs is at
intermidiate merging stage and the merger remnant  has not relaxed 
completely. On the other hand, the outer extension may be from
incompletely relaxed material or they can be stars formed
in secondary infall from tidally ejected material (e.g., Barnes 1988).

The five galaxies in class III show various degress of deviation from
the \r14 law, indeed \Ce\, is very well fitted by an exponential law.
Except \Ce, the other four ULIRGs are consistent
with their being at an early merging stage, 
both from their morphologies and surface brightness profiles (see \S 4.3).

The photometric properties and
the optical spectroscopic properties for our ULIRGs seem to be correlated.
It is interesting that the percentage of Seyfert 1s is very high (6/13)
in our sample compared with complete ULIRG samples, such as 
Lawrence et al (1999), Kim et al (1998), which has a percentage of Seyfert 1s
less than 15\%. All class II galaxies are Seyfert 1 galaxies/QSOs while two 
out of four class I galaxies are Seyfert 1/QSOs as well.
At least 5 of these 6 Seyfert 1s /or QSOs are strong or extremely
strong FeII emitters and most are NLS1 like galaxies.
This is very intriguing since our sample is solely selected
using the criterion
that there should be a single nucleus from HST snapshot images. 
In contrast, none of the Class III objects are Seyfert 1s and they are 
classified as either HIIs or LINERs, which
suggests that massive starbursts are present in these galaxies.

Let us now investigate the evolutionary tracks between these three
classes of ULIRGs. Morphologically speaking,
class I objects are clearly the most
relaxed, while class III objects in general show the most disturbed
morphologies and photometries, with class II objects somewhere in between.
This suggests an evolution sequence
from class III to class II, and then to class I, as relaxation proceeds
to ultimately form a relaxed elliptical galaxy or S0 galaxies,
depending on the initial mass ratios of the merging progenitors (Barnes 1998). 

In these three classes of galaxies, the interplay between star formation and
central AGN activities is undoubtedly very complex, nevertheless, here
we speculate qualitatively on the evolution of these galaxies based on
such an interplay. First, as galaxies begin to interact and merge,
massive nuclear starbursts are induced. In the mean time,
some gas also flows into the center to ignite
the central AGN activities, as indicated by numerical simulations.
The spectroscopic properties of galaxies at this stage
are a composite of the central AGN and circumnuclear starbursts, with
the latter being dominating.
Our class III objects are probably at this stage. As time goes on,
gas continuously fall into the center to 
fuel more intense central AGN activities, while the star formation has
subsided due to the gradual depletion of the gas supply.
Most of the dust is blown away, as a result, these galaxies are
mostly seen as Seyfert 1 galaxies. As discussed by Xia et al (1999),
the Seyfert 1 galaxy with strong FeII emission lines and with steep
soft X-ray photon index could be at such a transition stage,
namely these objects are emerging young QSOs. Our class II fits this
description qualitatively.
Finally, as relaxation proceeds even further, gas is mostly consumed
by star formation or blown out by superwind (Heckman et al 1990), and
the central AGN weakens and the merger remnant emerges as a
young elliptical galaxy -- our class I objects. If 
the central gas supply is not yet exhausted, then some of these galaxies
can also be Seyfert 1 galaxies (as two of our class I objects are).
This picture was first pointed out by Sanders et al (1988a) ten years ago.
Our observations seem to confirm this scenario.
Along this sequence, the gas behavior is also expected to
evolve. There is strong evidence that there are 
large (ten kpc scale), extended HI or emission-line nebula around some ULIRGs 
(Armus et al 1990; Hibbard \& Yun, 1997). The extended gas
distribution is formed by the secondary infall of gas thrown out during
the merging process. The infalling gas 
may form some outer extension of ULIRGs, as seen in our Class II
objects. It will be very interesting to obtain HI information
to study the evolution of gas in these galaxies (cf. Hibbard \& Van
Gorkom 1996).

To conclude, from our analysis of the photometry of ULIRGs, we found 
that these galaxies are excellent laboratories and the surface brightness
profile analysis is an important way
to study the merging proccess. The observational data provide a hint
that some of these galaxies reside in groups of galaxies and
multi-merger is not rare. Six out of these thirteen galaxies are 
Seyfert 1s or QSOs,
strongly suggesting that the AGN phenomenon is universal in the
merging process and is an integral part of elliptical galaxy formation.
A number of important works remain to be done. A multi-color
campaign of these objects will provide invaluable information about the
star formation sites and the relative importance of starbursts and central
nulcear activities. Furthermore, many galaxies 
(including a number of small and faint galaxies) are within the field of
the primary ULIRGs. Whether these galaxies are physically
associated with the ULIRGs awaits future spectroscopic observations. Such 
observations will provide insight not only on the environments of
ULIRGs but also on the nature of dwarf galaxies around these ULIRGs.

\begin{acknowledgements}
We thank Dr. H.J. Mo for useful discussions and the
BATC members for data reduction.
This project was partially supported by the NSF of China.
\end{acknowledgements}


\end{document}